     \documentclass[ twoside,journal]{IEEEtran}
    \usepackage{algorithm}
\usepackage{algorithmicx}
\usepackage{algpseudocode}
    \usepackage{makecell}
    \usepackage{array}
    \usepackage{graphicx,amssymb,amsmath}
    \usepackage{multicol}
    \usepackage[noadjust]{cite}
    \usepackage{setspace}
    \usepackage{subfigure}
    \usepackage{graphicx}
    \usepackage{float}
    \usepackage {url}
    \usepackage{stfloats}
    \usepackage{amsthm,pifont}
    \usepackage{flushend}
    \usepackage{cases,subeqnarray}
    \usepackage{bm,multirow,bigstrut}
    \usepackage{amsmath, amsthm, amssymb}
    \usepackage{textcomp}
    \usepackage{latexsym,bm}
    \usepackage{booktabs}
    \usepackage{xcolor}
    \usepackage{mathtools}
    \usepackage{dsfont}
    \usepackage{extarrows}
    \usepackage{epsfig}
    \usepackage{epsfig}
    \usepackage{epstopdf}
    \theoremstyle{plain}

    \theoremstyle{plain}

    \usepackage{amsmath}

\begin{document}
\title{Acceleration Estimation of Signal Propagation Path Length Changes for Wireless Sensing}
% Leveraging channel state information for fall detection in the indoor dynamic scenario

%

\author{
Jiacheng Wang,
Hongyang Du,
Dusit Niyato,~\IEEEmembership{Fellow,~IEEE}, Mu~Zhou, Jiawen~Kang, \\ and H. Vincent Poor,~\IEEEmembership{Life Fellow,~IEEE}% <-this % stops a spaceU
\thanks{J.~Wang, H.~Du and D. Niyato are with the School of Computer Science and Engineering, Nanyang Technological University, Singapore (e-mail: jiacheng.wang@ntu.edu.sg, hongyang001@e.ntu.edu.sg,  dniyato@ntu.edu.sg).}
\thanks{M.~Zhou is with School of Communication and Information Engineering, Chongqing University of Posts and Telecommunications, Chongqing, China (e-mail: zhoumu@cqupt.edu.cn).}
\thanks{J. Kang is with the School of Automation, Guangdong University of Technology, Guangzhou, China. (e-mail: kavinkang@gdut.edu.cn).}
\thanks{H. Vincent Poor is with the Department of Electrical and Computer Engineering, Princeton University, Princeton, NJ 08544 USA (e-mail: poor@princeton.edu).}
}
\maketitle
\begin{abstract}
As indoor applications grow in diversity, wireless sensing, vital in areas like localization and activity recognition, is attracting renewed interest. Indoor wireless sensing relies on signal processing, particularly channel state information (CSI) based signal parameter estimation. Nonetheless, regarding reflected signals induced by dynamic human targets, no satisfactory algorithm yet exists for estimating the acceleration of dynamic path length change (DPLC), which is crucial for various sensing tasks in this context. Hence, this paper proposes DP-AcE, a CSI based \underline{DP}LC \underline{ac}celeration \underline{e}stimation algorithm. We first model the relationship between the phase difference of adjacent CSI measurements and the DPLC's acceleration. Unlike existing works assuming constant velocity, DP-AcE considers both velocity and acceleration, yielding a more accurate and objective representation. Using this relationship, an algorithm combining scaling with Fourier transform is proposed to realize acceleration estimation. We evaluate DP-AcE via the acceleration estimation and acceleration-based fall detection with the collected CSI. Experimental results reveal that, using distance as the metric, DP-AcE achieves a median acceleration estimation percentage error of 4.38\%. Furthermore, in multi-target scenarios, the fall detection achieves an average true positive rate of 89.56\% and a false positive rate of 11.78\%, demonstrating its importance in enhancing indoor wireless sensing capabilities.

%{Indoor integrated sensing and communication (ISAC) relies on electromagnetic signal processing, and using channel state information (CSI) to estimate signal parameters is an essential component of this process. However, no satisfactory algorithm yet exists for estimating dynamic signal propagation path length change (DPLC) acceleration, which is crucial for sensing tasks like passive human tracking and activity recognition. Therefore, in this paper we propose DP-AcE, a CSI based \underline{DP}LC \underline{Ac}celeration \underline{E}stimation procedure. We first model the relationship between the phase difference of adjacent CSI measurements and the DPLC acceleration. Unlike existing works that assume constant velocity, DP-AcE consider both velocity and acceleration, resulting in a more objective and accurate representation. Based on this relationship, an algorithm combining two-dimensional Fourier transform with scaling is proposed to estimate DPLC acceleration. We evaluate the proposed method using DP-AcE-based fall detection as an example with CSI collected from commercial wireless devices. Experimental results reveal that, using distance as the metric, DP-AcE achieves a median estimation percentage error of 4.38\% for acceleration estimation. Furthermore, in multi-occupant scenarios, it realizes an average fall detection accuracy of 89.56\% and a false alarm rate of 11.78\%, demonstrating its importance in further enhancing indoor wireless sensing capabilities.}
\end{abstract}

\begin{IEEEkeywords}
Wireless sensing, channel state information, signal parameter estimation.
\end{IEEEkeywords}
\IEEEpeerreviewmaketitle
\section{Introduction}
    With the rapid proliferation of smart devices and the advancement of wireless technologies, such as integrated sensing and communication (ISAC)~\cite{liu2022survey, yang2022secure}, wireless sensing has become one of the pivotal techniques for human-centered computing applications. In indoor scenarios, wireless sensing utilizes diverse algorithms to extract time, frequency, and spatial signal features. These are then analyzed by using methods like statistical analysis and machine learning to perform sensing tasks~\cite{wang2023unified}, such as indoor localization~\cite{wang2023through}, tracking~\cite{qian2018widar2}, and behavior recognition~\cite{wang2015understanding}. Given the pervasive nature of wireless signals, wireless sensing exhibits a broad operational range~\cite{zeng2021exploring}. Moreover, with its ability to operate in both active and passive modes, it enjoys high acceptance among users, demonstrating substantial market potential.

    As a vital component of indoor wireless sensing, WiFi channel state information (CSI) processing has emerged as an important research focus, attracting extensive exploration in recent times. For instance, in~\cite{zhang20223d}, the authors consider both occlusion and perturbation effects caused by electromagnetic wave propagation and propose a CSI-based computational imaging approach for 3D environment sensing. In~\cite{kotaru2015spotfi}, the authors employ a two-dimensional multiple signal classification algorithm (2D-MUSIC) to estimate the angle of arrival (AoA) and time of flight (ToF) of wireless signals based on CSI, so as to achieve user localization. MUSIC is a subspace-based super-resolution estimation algorithm, which leverages the orthogonality between a steering matrix and noise subspace, attained through matrix decomposition, to estimate parameters via an iterative search~\cite{gupta2015music}. Building on MUSIC, the authors in~\cite{li2017indotrack} present a novel algorithm to estimate the velocity of the dynamic path length change (DPLC)\footnote{The dynamic path refers to the propagation path of the signal introduced by the dynamic human target. The length of this path changes with the movement of the human body. Hence, they refer to this change as DPLC.}. This approach assume that the velocity of DPLC is a constant over a short period of time and remodels the steering matrix to perform the estimation. Furthermore, in~\cite{hu2021defall}, the authors apply the statistical theory of electromagnetic waves to model the relationship between the auto-correlation function (ACF) of CSI and the velocity of moving human. Then, the estimation of a moving human velocity is achieved by detecting the peak points of the ACF.

    The diverse parameters estimated from CSI can be harnessed for various sensing tasks. For example, the frequency components within CSI can be exploited for human activity recognition~\cite{yousefi2017survey} and behavior analysis~\cite{wang2023guiding}, while the AoA and ToF of the signal can be utilized for localization~\cite{wang2023through}. Among these parameters, the acceleration of DPLC is indispensable~\cite{zeng2018fullbreathe}, since it reflects the distinct dynamic characteristics of different propagation paths, enabling differentiation and motion state analysis of multiple human targets. This can provide support for diverse wireless sensing applications, such as human detection, tracking, fall detection, breath detection~\cite{hu2020wifi, zeng2018fullbreathe}, as well as the walking gait monitoring~\cite{mei2019determination}. While significant, a satisfactory estimation approach for this pivotal parameter remains elusive. We posit that the fundamental reasons for this impasse primarily stem from the following two aspects:
    \begin{itemize}
        \item[{\textbf{A1)}}] From the modeling perspective, most existing works~\cite{li2017indotrack,qian2017widar,tian2023localization} assume that the velocity of DPLC over a short period of time is constant. While this holds to a certain extent, most human movements are not essentially uniform~\cite{umberger2010stance}\footnote{Studies have shown that human walking can be divided into a stance phase and a swing phase, and each phase includes a double limb support (DLS) and single limb support (SLS) stages. The switching between these two stages causes a change in acceleration. Therefore, the acceleration during walking is not zero. This is explained in more detail in Section III.}. For example, the walking process of a human is a motion with non-zero acceleration, indicating that the velocity of DPLC changes over time, rather than being a constant value. Therefore, the signal model formulated under the aforementioned assumption lacks accuracy and objectivity.
       \item[{\textbf{A2)}}] From the algorithmic perspective, the direct estimation of the acceleration from CSI is challenging, as the phase accumulation components caused by velocity and acceleration are coupled, resulting in nonlinear phase changes. For instance, the MUSIC algorithm used in~\cite{li2017indotrack} and~\cite{tian2023localization} can only be applied for velocity estimation when the phase changes linearly between CSI measurements. Some studies~\cite{zhang2018wispeed} use electromagnetic theory to obtain the acceleration of DPLC indirectly by estimating the velocity of the moving target. However, they struggle to estimate the acceleration of different targets in multi-target cases.

   \end{itemize}

   To address these challenges, we propose a novel CSI based DPLC acceleration estimation (DP-AcE) algorithm. Specifically, DP-AcE models the relationship between the phase difference of adjacent CSI measurements and the velocity and acceleration of DPLC. Based on this relationship, an algorithm combining scaling with the two-dimensional Fourier transform (2D-FT) is proposed to estimate the acceleration of DPLC. DP-AcE distinguishes itself from previous methods by incorporating acceleration into the modeling process, leading to a more accurate and objective signal model. On this basis, it can directly estimate the acceleration of DPLC from CSI, without the need for any searching operations. We evaluate DP-AcE from two aspects, including acceleration estimation accuracy and its corresponding application, using fall detection an example. Our results demonstrate that the proposed algorithm can effectively estimate the acceleration of DPLC with higher accuracy compared to existing algorithms. Furthermore, DP-AcE based fall detection can operate in scenarios with multiple human targets, without the need for re-training. In summary, the main contributions of this work are as follows:

\begin{itemize}
\item We propose a model that establishes the relationship between the phase change of two adjacent CSI measurements and the velocity and acceleration of DPLC. In contrast to prior studies that solely considers the velocity, our model exhibits greater objectivity and accuracy (to address \textbf{A1}).

\item We propose a novel algorithm that combines scaling with the 2D-FFT to directly estimate the acceleration of DPLC based on the constructed model. Given that each path has a unique acceleration, the proposed algorithm can discriminate different dynamic propagation paths (to address \textbf{A2}).

\item The experimental results show that using distance as the metric, DP-AcE achieves a median percentage error of 4.38\% for acceleration estimation, outperforming existing methods. Moreover, the DP-AcE based fall detection system achieves an average true positive rate of 89.56\% and a false positive rate of 11.78\%, when multiple dynamic targets appear.
\end{itemize}

The remainder of this paper is organized as follows. In Section II, we review some related works. In Section III, we present the detailed design of our system, including the signal model, estimation algorithm, and fall detection scheme. Section IV presents our experimental evaluation, followed by a conclusion in Section V.
\section{related work}
In this section, we focus on analyzing the literature pertaining to activity detection and recognition using wireless signals.

\subsection{Human Activity Recognition}
Based on the relationship between activity characteristics and location, the authors in~\cite{wang2014eyes} develop a location-oriented activity recognition system, which can distinguish between different in-place activities, such as washing dishes and walking. Wei $et\ al.$~\cite{wei2015radio} consider the impact of radio frequency interference on CSI measurements and use several countermeasures to mitigate this impact to further improve the activity recognition performance. Unlike~\cite{wang2014eyes} and~\cite{wei2015radio}, Li $et\ al.$~\cite{li2020location} propose a location-free activity recognition system, which first estimates the angle difference of arrival and then uses a trained Bidirectional Long Short-Term Memory network to complete recognition. In CARM~\cite{wang2017device}, the authors propose a CSI-speed model to quantify the relation between CSI dynamics and human movement speeds. Then, they build a CSI-activity model to quantify the relation between human movement speeds and activities, so as to recognize daily activities via a Hidden Markov Model. Besides activities, CSI can also be used for subtle motion detection. For example, WiCatch~\cite{tian2018wicatch} estimates and tracks the relative AoA of hand-induced reflection and accomplishes gesture recognition via a trained support vector machine (SVM) classifier. PhaseBeat~\cite{wang2017phasebeat} provides an analysis of the CSI phase difference data with respect to its stability and periodicity, and uses root-MUSIC and fast Fourier transform (FFT) to estimate the breathing frequencies and heart rates of multiple individuals.

\subsection{Human Fall Detection}
The first fall detection work using commodity WiFi devices is believed to be WiFall~\cite{wang2016wifall}. WiFall first extracts signal features, such as normalized standard deviation and signal entropy, from CSI amplitude, through CSI aggregation, outlier removal, and singular value decomposition. Then, the SVM and Random Forests classifiers are trained to classify different human activities and achieve fall detection. Unlike WiFall, RTFall~\cite{wang2016rt} uses the CSI phase difference over two antennas and determines the sharp power profile decline pattern of the fall in the time-frequency domain. Leveraging these observations, RTFall segments the CSI data stream and extracts features, such as median absolute deviation and signal change velocity, to train the SVM classifier for fall detection. Different from the above works, which only considers features in the time domain, FallDeFi~\cite{palipana2018falldefi} uses the short-time Fourier transform (STFT) to extract time-frequency features and the sequential forward selection algorithm to single out features that are resilient to changes in the environment. Then, these features are used to train a classifier for fall detection. Recent works Wispeed~\cite{zhang2018wispeed} and Defall~\cite{hu2020wifi} establish a link between the auto-correlation function of the CSI and the speed of a moving human. Through this, they can estimate the velocity of the moving human and accomplish fall detection by analyzing the change in the estimated velocity.

Different from the previous work, in this paper, we establish a more accurate relationship between the phase accumulation between two adjacent CSI data packets and the velocity and acceleration of DPLC. On this basis, we propose a novel algorithm to estimate velocity and acceleration from CSI directly. At last, we evaluate the acceleration estimation performance and use the fall detection as an example to illustrate the importance of acceleration estimation in wireless sensing.
\section{SYSTEM DESIGN}
\begin{figure*}[htp]
    \centering
    \includegraphics[width=1\textwidth]{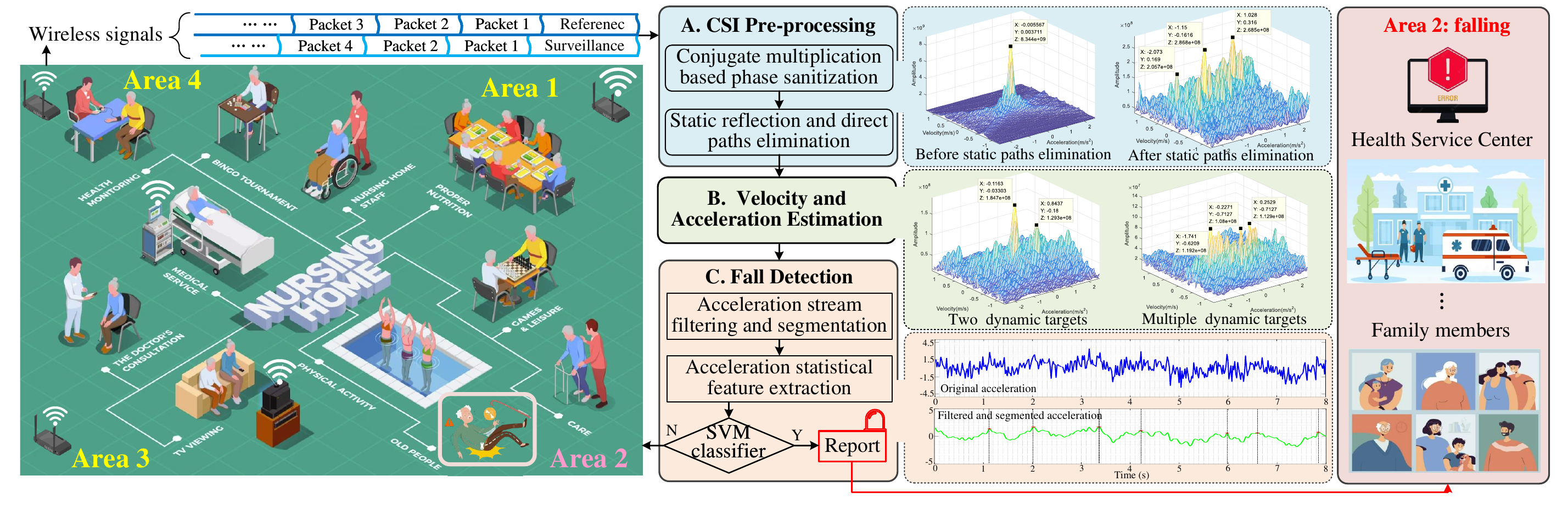}
    \caption{The overall framework of the proposed DP-AcE. Generally, the system includes three parts, i.e., A. CSI Pre-processing, B. Velocity and Acceleration Estimation, and C. Fall Detection, represented by blue, green, and red colors, respectively. When a fall is detected, the system reports immediately; otherwise, it continues to acquire data for processing. The algorithm proposed in this paper can distinguish the acceleration corresponding to multiple human targets, making DP-AcE the first algorithm suitable for scenarios with multiple human targets.}
    \label{framework9}
    \end{figure*}

This section presents the detailed design of the proposed system, which can be roughly divided into three parts, including CSI pre-processing, acceleration estimation, and fall detection. Taking the nursing home an example, the overall framework of the proposed system is shown in Fig.~\ref{framework9}.
\subsection{CSI Pre-processing}
\begin{figure}[tbp!]
  \centering
  \includegraphics[height=2.2cm]{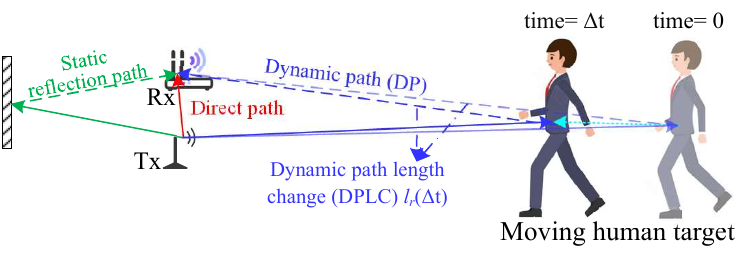}
  \caption{The process of signal propagation involves multiple paths. The static reflection path corresponds to the signal reflected by the static object, while the direct path corresponds to the signal traveling from the transmitter to the receiver directly. The dynamic path (DP) refers to the propagation path of the signal introduced by the moving human target. As the human moves, the length of the DP changes accordingly. This change is referred to as dynamic path length change (DPLC).}
  \label{PPGT}
\end{figure}

Assuming that a transmission pair based on the IEEE 802.11 protocol is deployed to monitor a moving human, as shown in Fig.~\ref{PPGT}. The transmitter (Tx) sends an orthogonal frequency-division multiplexed (OFDM) signal, and the receiver (Rx) with $M$ antennas is employed to receive signals.
To eliminate phase errors (which will be explained later in this section), the first antenna of the Rx is used to capture the direct path signal and we treat it as a reference signal. Meanwhile, the remaining antennas are employed to receive the multipath signals, which are considered as surveillance signals. The Rx adopts training sequences to synchronize and estimate the channel parameters. Thereby, the CSI of the reference signal extracted from the $k$-th subcarrier can be denoted as
\begin{align}\label{eq1}
\begin{array}{l}
{H_{1,k}} = \left[ {\sum\limits_{d = 1}^D {\alpha _{1,k}^d\exp \left( { - j2\pi {f_k}\tau _1^d} \right)} } \right. \\
\left. {{\rm{            }}\sum\limits_{r = 1}^R {\alpha _{1,k}^r\exp \left( { - j2\pi {f_k}\tau _1^r} \right)} } \right] \times \exp \left( { - j2\pi \varepsilon } \right) + {n_{1,k}},
\end{array}
\end{align}
where $D$ and $R$ represent the numbers of direct and reflected propagation paths\footnote{In practice, these two types of signals are indistinguishable. Here, they are divided into two categories for the purpose of facilitating understanding of the differences in signal strength.}, respectively, ${\alpha _{1,k}^d}$ and ${\alpha _{1,k}^r}$ are complex attenuation factors, ${f_k}$ is the frequency of the $k$-th subcarrier, ${\tau _1^d}$ is the ToF of the direct path signals, ${\tau _1^r}$ is the ToF of the reflected signal introduced by objects and human targets, ${\varepsilon}$ is the time offset introduced by the packet detection delay (PDD), sampling frequency offset (SFO), and central frequency offset (CFO), and ${n_{1,k}}$ is noise. Here, the main phase error can be specifically expressed as $\varepsilon  = {f_k}\left( {{\varepsilon _{PDD}} + {\varepsilon _{SFO}}} \right) + {\varepsilon _{CFO}}$. The term ${\varepsilon _{PDD}}$ is a random error, which is introduced by the inability of packet detectors to accurately detect the arrival of data packets. The ${\varepsilon _{SFO}}$ is introduced due to different sampling rates between the Tx and Rx, which remains relatively stable over short periods of time. The lack of perfect synchronization between the Tx and Rx results in the introduction of ${\varepsilon _{CFO}}$. Although the corrector compensates for the CFO, due to hardware imperfections, the compensation is incomplete, leading to residual errors contained in the CSI. Similarly, the CSI of surveillance signal extracted from the $k$-th subcarrier of the $m$-th antenna is
\begin{align}\label{eq2}
\begin{array}{l}
{H_{m,k}}{\rm{ = }}\left[ {\sum\limits_{d' = 1}^{D'} {\alpha _{m,k}^{d'}\exp \left( { - j2\pi {f_k}\tau _m^{d'}} \right)} } \right. \\
\left. {\sum\limits_{r' = 1}^{R'} {\alpha _{m,k}^{r'}\exp \left( { - j2\pi {f_k}\tau _m^{r'}} \right)} } \right] \times \exp ( - j2\pi \varepsilon ) + {n_{m,k}},
\end{array}
\end{align}
where $m = 2,{\ } \ldots {{, }\ }M$, $D'$ and $R'$ are the numbers of direct and reflection propagation paths, ${\alpha _{m,k}^{d'}}$ and ${\alpha _{m,k}^{r'}}$ are the complex attenuation factors, $\tau _{m}^{d'}$ and $\tau _{m}^{r'}$ are the propagation delay, respectively, and ${n_{m,k}}$ is noise. The receiver processes the reference signal and surveillance signal based on the clock generated by the same oscillator, and therefore, the phase errors contained in both signals are the same. Based on this observation, conjugate multiplication is conducted between two channels to remove the phase errors
\begin{align}\label{eq3}
{H_{m1,k}} &= {H_{m,k}} \times {{\bar H}_{1,k}}\\ \notag
 &= \underbrace {\sum\limits_{d' = 1}^{DD'} {\alpha _{m1,k}^{d'}} \exp \left[ { - j2\pi {f_k}\left( {\tau _m^{d'} - \tau _1^d} \right)} \right]}_{direct\;component}\\ \notag
 &+ \underbrace {\sum\limits_{dr' = 1}^{DR'} {\alpha _{1m,k}^{dr'}\exp \left[ { - j2\pi {f_k}\left( {\tau _m^{r'} - \tau _1^d} \right)} \right]} }_{cross\;component{\ { 1}}}\\ \notag
& + \underbrace {\sum\limits_{d'r = 1}^{D'R} {\alpha _{m1,k}^{d'r}\exp \left[ { - j2\pi {f_k}\left( {\tau _m^{d'} - \tau _1^r} \right)} \right]} }_{cross\;component{\ { 2}}}\\ \notag
&{\rm{ + }}\underbrace {\sum\limits_{r' = 1}^{RR'} {\alpha _{m1,k}^{r'}\exp \left[ { - j2\pi {f_k}\left( {\tau _m^{r'} - \tau _1^r} \right]} \right)} }_{reflection\;component}\\ \notag
&+ \Gamma  + \Psi  + \underbrace {{n_{m,k}} \times {n_{1,k}}}_{noise} ,
\end{align}
where ${\bar H_{1,k}}$ is the complex conjugate of ${H_{1,{\rm{ }}k}}$, $\alpha _{m1,k}^{d'}{\rm{ = }}\alpha _{1,k}^d \times \alpha _{m,k}^{d'}$, $\alpha _{m1,k}^{r'}{\rm{ = }}\alpha _{1,k}^r \times \alpha _{m,k}^{r'}$, $\alpha _{1m,k}^{dr'}{\rm{ = }}\alpha _{1,k}^d \times \alpha _{m,k}^{r'}$, $\alpha _{m1,k}^{d'r}{\rm{ = }}\alpha _{m,k}^{d'} \times \alpha _{1,k}^r$, $\Gamma $ is the product of the CSI in ${H_{m,k}}$ and the noise in ${{\bar H}_{1,k}}$, and $\Psi $ is the product of the CSI in ${{\bar H}_{1,k}}$ and the noise in ${H_{m,k}}$. After conjugate multiplication, three pivotal points can be observed.
\begin{itemize}
\item First, the first four components in~\eqref{eq3} no longer include phase errors. As a result, it is possible to estimate the signal parameters based on ${H_{m1,k}}$, like signal ToF and Doppler, and theoretically, the estimation results will not be affected by phase errors.

\item Second, ${\tau_1^d}$ refers to the ToF of the direct path signal. In the case where the positions of the transceiver devices remain unchanged, ${\tau_1^d}$ is a constant that can be measured and compensated.

\item Third, the direct component has the strongest signal energy, followed by the cross components, then the reflection component, and the rest. The DPs are included in all components except for the noise and the first component. Among them, the cross terms hold the strongest energy without phase errors, thereby providing an opportunity for extracting acceleration of DPLC.
\end{itemize}

\begin{figure}[tbp!]
  \centering
  \includegraphics[height=6.5cm]{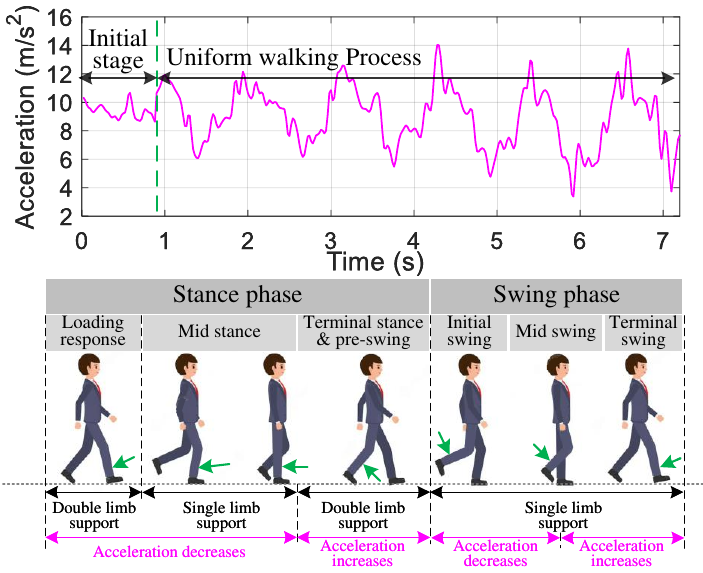}
  \caption{ The acceleration captured by the inertial measurement unit (IMU) and the human walking process. The walking of a human consists of a stance phase and a swing phase. Taking the marked right foot as an example, the stance phase contains two double limb support (DLS) stages and one single limb support (SLS) stage. During the stance stage, the human's center of gravity moves upward with the transition from the DLS to SLS, causing a decrease in forward acceleration. During the second DLS, the body's center of gravity descends with the transition from SLS back to DLS, leading to an increase in acceleration.}
  \label{MEMSD}
\end{figure}

Without loss of generality, we consider that the sanitized CSI presented in~\eqref{eq3} is obtained at time 0, and the $r$-th propagation path is the DP. As the human walks in the monitored area, the propagation length of the DP changes accordingly. Assume the human target walks towards the transceiver pair as shown in Fig.~\ref{PPGT}. According to the existing research~\cite{umberger2010stance}, this walking process can be divided into the stance and swing phases and each phase contains the double limb support (DLS) stage and single limb support (SLS) stage, as shown in the Fig.~\ref{MEMSD}. The acceleration of the human body changes with the switching between the DLS stage and the SLS stage~\cite{yang2015three}. We employ the nine-axis IMU to measure acceleration during human walking. The results, presented in Fig.~\ref{MEMSD}, corroborate the conclusions of existing research.
Hence, we assume that the human target walks towards the transceiver pair with the velocity of ${v_r}$ and acceleration of ${a_r}$, as shown in Fig.~\ref{PPGT}. Then, after a sampling interval time $\Delta t$, the DPLC is
\begin{equation}\label{eq4}
{l_r}\left( {\Delta t} \right) = {v_r}\Delta t + \frac{1}{2}{a_r}{\left( {\Delta t} \right)^2}.
\end{equation}
This path length change yields a phase shift to the CSI of the $k$-th subcarrier corresponding to the DP, which is
\begin{align}\label{eq5}
{P_{k,r}}(\Delta t) &= \exp \left[ { - j2\pi  {f_k}\frac{{2{l_r}\left( {\Delta t} \right)}}{c}} \right] \\ \notag
&= \exp \left[ { - j2\pi \frac{{{f_k}}}{c}2{v_r}\Delta t - j2\pi \frac{{{f_k}}}{c}{a_r}{{\left( {\Delta t} \right)}^2}} \right],
\end{align}
where $c$ is the propagation speed of electromagnetic wave. As different reflection paths have different accelerations\footnote{For some static reflection paths introduced by objects, such as walls, furniture with metal shells, and electrical appliances, the corresponding velocity and acceleration are zero. These reflections will be filtered out together with the direct path when performing zero frequency filtering later.}, at the time $\Delta t$, the cross component in ${H_{m1,k}}$ can be denoted as
\begin{align}\label{eq6}
{{H'}_{m1,k}} &=\underbrace {\sum\limits_{dr' = 1}^{DR'} {\alpha _{1m,k}^{dr'}\exp \left[ { - j2\pi {f_k}\left( {\tau _m^{r'} - \tau _1^d} \right)} \right]{P_{k,dr'}}} }_{cross\;component{\ { 1}}} \\ \notag
&+ \underbrace {\sum\limits_{d'r = 1}^{D'R} {\alpha _{m1,k}^{d'r}\exp \left[ { - j2\pi {f_k}\left( {\tau _m^{d'} - \tau _1^r} \right)} \right]{P_{k,d'r}}} }_{cross\;component{\ { 2}}}.
\end{align}

From \eqref{eq4} to \eqref{eq6}, we can see that the DPLC over time is directly related to the velocity and acceleration of the moving human, and ${\Delta t}$, which distinguishes it from the other propagation paths. However, as discussed previously, the strength of the DP is lower than that of the direct path signal, which can cause interference during the estimation process. Furthermore, the static reflection path signal, which may also be stronger than DP, aggravating such interference. Therefore, it is crucial to eliminate this interference before estimation. As the phase of these propagation paths does not change over time, indicating that they have a frequency of zero, it is natural to consider eliminating them in the frequency domain. Concretely, we employ a window of length $W$ to extract data from the CSI stream and construct the following matrix
\begin{align}\label{e7}
{{\bf{H}}_{m1,k}} &= \left[ {{H_{m1,k}}\left( 0 \right),{\ } \ldots }, \right.\left. {{\ } {H_{m1,k}}\left( {(W - 1)\Delta t} \right)} \right].
\end{align}
Then, the fast Fourier transformation (FFT) is used to convert ${{\bf{H}}_{m1,k}}$ into frequency domain
\begin{align}\label{eq8}
{{{\bf{H'}}_{m1,k}}} =  {{\mathcal F}}({{\bf{H}}_{m1,k}}),
\end{align}
where ${\mathcal F}\left(  \cdot  \right)$ is the FFT operator. On this basis, we set the value of frequency zero in ${{\bf{H'}}_{m1,k}}$ to zero, so as to eliminate the interference. After that, ${{{\bf{H'}}_{m1,k}}}$ is converted back to the time domain via the inverse fast Fourier transform (IFFT). Here, the matrix converted back to the time domain is denoted as ${{\bf{J}}_{m,k}}$, which no longer contains the CSI corresponding to the signals along the direct path and static reflection path.

\subsection{Velocity and Acceleration Estimation}
Upon completion of CSI pre-processing, the ${{\bf{J}}_{m,k}}$ is utilized to estimate the velocity and acceleration. To facilitate understanding, we assume that ${{\bf{J}}_{m,k}}$ only contains the cross component 2 in~(\ref{eq3})\footnote{The other components can be derived in the same way.} and simplify $\alpha _{m1,k}^{d'r}$ to $\alpha _{m,k}^{r}$. On this basis, we calculate the parametric symmetric instantaneous auto-correlation (PSIA) function of ${{\bf{J}}_{m,k}}$ as
\begin{align}\label{eq9}
A_J^C &\left( {\Delta t,\tau } \right) = {J_{m,k}}\left( {\Delta t + \frac{{\tau  + {t_d}}}{2}} \right){{\bar J}_{m,k}}\left( {\Delta t - \frac{{\tau  + {t_d}}}{2}} \right)\notag\\
 &= \underbrace {\sum\limits_{r = 1}^{R - 1} {\sum\limits_{i = r + 1}^R {\left[ {A_{J_{m,k}^{\left( r \right)}J_{m,k}^{\left( i \right)}}^C\left( {\Delta t,\tau } \right) + A_{J_{m,k}^{\left( i \right)}J_{m,k}^{\left( r \right)}}^C\left( {\Delta t,\tau } \right)} \right]} } }_{cross{\ }terms}\notag\\
 &+ \underbrace {\sum\limits_{r = 1}^R {{{\left( {\alpha _{m,k}^{r}} \right)}^2}\exp \left[ { - j4\pi {f_k}\left( {{v_r} + {a_r}\Delta t} \right)\frac{{\left( {\tau  + {t_d}} \right)}}{c}} \right]} }_{auto{\ }terms}\notag\\
 &= \sum\limits_{r = 1}^{R - 1} {\sum\limits_{i = r + 1}^R {\left[ {A_{J_{m,k}^{\left( r \right)}J_{m,k}^{\left( i \right)}}^C\left( {\Delta t,\tau } \right) + A_{J_{m,k}^{\left( i \right)}J_{m,k}^{\left( r \right)}}^C\left( {\Delta t,\tau } \right)} \right]} }\notag\\
 &+ \sum\limits_{r = 1}^R {A_{J_{m,k}^{\left( r \right)}}^C\left( {\Delta t,\tau } \right),}
\end{align}
where ${t_d}$ is a constant time delay related to a scaling operator, $A_{J_{m,k}^{\left( r \right)}J_{m,k}^{\left( i \right)}}^C$ and $A_{J_{m,k}^{\left( i \right)}J_{m,k}^{\left( r \right)}}^C$ are the cross terms, and $A_{J_{m,k}^{\left( r \right)}}^C$ is the auto term. From ~\eqref{eq9}, it can be seen that $\Delta t$ and variable $\tau $ are coupled together in terms of exponential phase, which can blur the estimation result. Motivated by the keystone transformation~\cite{lv2011lv}, a scaling operator is defined to solve the coupling problem between $\Delta t$ and $\tau $. Concretely, for any given function $F$ related to $\left( {\Delta t,\tau } \right)$, the specific scaling operation is defined as
\begin{align}\label{eq10}
Z\left[ {F\left( {\Delta t,\tau } \right)} \right] = F\left[ {{t \mathord{\left/
 {\vphantom {t {s\left( {\tau  + {t_d}} \right)}}} \right.
 \kern-\nulldelimiterspace} {s\left( {\tau  + {t_d}} \right)}},\tau } \right],
\end{align}
where $t$ is the scaled time and $s$ is the scaling factor. Thus, conducting the scaling operation on $A_J^C\left( {\Delta t,\tau } \right)$, we have
\begin{align}\label{eq11}
Z&\left[ {A_J^C\left( {\Delta t,\tau } \right)} \right] = \sum\limits_{r = 1}^R {\left( {{{\left( {\alpha _{m,k}^{r}} \right)}^2}} \right.} \\ \notag
&\left. { \times \exp \left[ { - j4\pi {f_k}{v_r}\frac{{\left( {\tau  + {t_d}} \right)}}{c}} \right]\exp \left( { - j4\pi {f_k}{a_r}\frac{t}{{sc}}} \right)} \right)\\ \notag
 &+ \sum\limits_{r = 1}^{R - 1} {\sum\limits_{i = r + 1}^R {Z {\left( \left[{A_{J_{m,k}^{\left( r \right)}J_{m,k}^{\left( i \right)}}^C\left( {t,\tau } \right) + A_{J_{m,k}^{\left( i \right)}J_{m,k}^{\left( r \right)}}^C\left( {t,\tau } \right)} \right]\right)} } } ,
\end{align}
where
\begin{align}\label{eq12}
&Z  {\left( \left[ {A_{J_{m,k}^{\left( r \right)}J_{m,k}^{\left( i \right)}}^C\left( { t,\tau } \right) + A_{J_{m,k}^{\left( i \right)}J_{m,k}^{\left( r \right)}}^C\left( { t,\tau } \right)}\right] \right)} \\ \notag
  &= 2\alpha _{m,k}^r\alpha _{m,k}^i \times {\mathop{\rm Re}\nolimits} \left\{ {\exp \left[ { - j2\pi {f_k}\frac{{4t\left( {{v_r} - {v_i}} \right)}}{{cs\left( {\tau  + {t_d}} \right)}}} \right]} \right.\\ \notag
&\left. { - j2\pi {f_k}\frac{2}{c}\left[ {\frac{{\left( {{a_r} - {a_i}} \right){t^2}}}{{{s^2}{{\left( {\tau  + {t_d}} \right)}^2}}} + \frac{{\left( {{a_r} - {a_i}} \right){{\left( {\tau  + {t_d}} \right)}^2}}}{4}} \right]} \right\}\\ \notag
 & \times \exp \left( { - j2\pi {f_k}\frac{2}{c}\left[ {\left( {{v_r} + {v_i}} \right)\left( {\tau  + {t_d}} \right) - \left( {{a_r} + {a_i}} \right)\frac{t}{s}} \right]} \right)  \\ \notag
 &= 2\alpha _{m,k}^{\left( r \right)}\alpha _{m,k}^{\left( i \right)}  A_{J_{m,k}^{\left( r \right)}J_{m,k}^{\left( i \right)}}^{C1}\left( { t,\tau } \right) {\mathop{\rm Re}\nolimits} \left[ {A_{J_{m,k}^{\left( r \right)}J_{m,k}^{\left( i \right)}}^{C2}\left( { t,\tau } \right)} \right],
\end{align}
and ${\mathop{\rm Re}\nolimits} \left(  \cdot  \right)$ indicates the operation of taking the real part. As can be seen from~\eqref{eq11} and \eqref{eq12}, after the scaling operation, the coupling problem has been settled. On this basis, the two-dimensional (2-D) FFT is conducted on $Z\left[ {A_J^C\left( {\Delta t,\tau } \right)} \right]$ with respect to $\tau$ and $t$ to generate the velocity-acceleration (V-A) plane
\begin{align}\label{eq13}
{G_J}\left( {v,a} \right) &={\mathcal{{ F}}_\tau }\left\{ {{\mathcal{{ F}}_t}\left[ {Z\left( {A_J^C\left( {\Delta t,\tau } \right)} \right)} \right]} \right\} \\ \notag
&= \underbrace {\sum\limits_{r = 1}^R {{G_{J_{m,k}^{\left( r \right)}}}\left( {v,a} \right)} }_{auto{\ }terms} \\ \notag
&+ \underbrace {\sum\limits_{r = 1}^{R - 1} {\sum\limits_{i = r + 1}^R {{G_{J_{m,k}^{\left( r \right)}J_{m,k}^{\left( i \right)}}}\left( {v,a} \right)} } }_{cross{\ }terms}.
\end{align}

Specifically, for the auto terms, the analysis of the transformation process and corresponding results of 2D-FFT are as follows. Performing the FFT on the $r$-th auto term in~\eqref{eq13} with respect to $t$ we have
\begin{align}\label{eq14}
{G_{J_{m,k}^{\left( r \right)}}}\left( {\tau ,a} \right) &= {{\cal F}_t}\left( {{{\left( {\alpha _{m,k}^{\left( r \right)}} \right)}^2}} \right. \times \exp \left( { - j4\pi {f_k}{a_r}\frac{t}{{sc}}} \right) \\ \notag
&\left. { \times \exp \left[ { - j4\pi {f_k}{v_r}\frac{{\left( {\tau  + {t_d}} \right)}}{c}} \right]} \right) \\ \notag
 &= {\left( {\alpha _{m,k}^{\left( r \right)}} \right)^2}\exp \left[ { - j4\pi {f_k}{v_r}\frac{{\left( {\tau  + {t_d}} \right)}}{c}} \right] \\ \notag
 &\times \int_{ - \infty }^\infty  {\exp \left[ {j4\pi {f_k}\frac{2}{c}\left( {a - \frac{{{a_r}}}{s}} \right)t} \right]dt} \\ \notag
 &= {\left( {\alpha _{m,k}^{\left( r \right)}} \right)^2}\exp \left( { - j4\pi {f_k}{v_r}\frac{{\tau  + {t_d}}}{c}} \right) \\ \notag
 &\times \delta \left( {a - \frac{{{a_r}}}{s}} \right),
\end{align}
where $\delta \left(  \cdot  \right)$ represents the Dirac delta function. After that, the FFT with respect to $\tau$ is conducted on ${G_{J_{m,k}^{\left( r \right)}}}\left( {\tau ,a} \right)$ leads to
\begin{align}\label{eq15}
{G_{J_{m,k}^{\left( r \right)}}}\left( {v,a} \right) &= {{\cal F}_\tau }\left( {{{\left( {\alpha _{m,k}^{\left( r \right)}} \right)}^2}\delta \left( {a - \frac{{{a_r}}}{s}} \right)} \right.\\ \notag
&\left. { \times \exp \left[{ - j4\pi {f_k}{v_r}\frac{\left( {\tau  + {t_d}} \right) }{c}} \right]} \right)\\ \notag
 &= {\left( {\alpha _{m,k}^{\left( r \right)}} \right)^2}\delta \left( {v - {v_r}} \right)\delta \left( {a - \frac{{{a_r}}}{s}} \right)\\ \notag
 &\times \exp \left( { - j4\pi {f_k}{v_r}\frac{{{t_d}}}{c}} \right).
\end{align}

From the result presented in~\eqref{eq15}, we can see that each auto term generates an impulse in V-A plane. In a similar way, the transformation result of the cross term can be also obtained. To facilitate the derivation, we assume that $\Delta {v_{ri}} = \left( {{{ - {f_k}} \mathord{\left/
 {\vphantom {{ - {f_k}} c}} \right.
 \kern-\nulldelimiterspace} c}} \right)\left( {{v_r} - {v_i}} \right)$, $\nabla {v_{ri}} = \left( {{{ - {f_k}} \mathord{\left/
 {\vphantom {{ - {f_k}} c}} \right.
 \kern-\nulldelimiterspace} c}} \right)\left( {{v_r} + {v_i}} \right)$, $\tilde v = \left( {v - {{\left( {{v_r} + {v_i}} \right)} \mathord{\left/
 {\vphantom {{\left( {{v_r} + {v_i}} \right)} 2}} \right.
 \kern-\nulldelimiterspace} 2}} \right)
$, $\Delta {a_{ri}} = \left( {{{{f_k}} \mathord{\left/
 {\vphantom {{{f_k}} c}} \right.
 \kern-\nulldelimiterspace} c}} \right)\left( {{a_r} - {a_i}} \right)$, $\nabla {a_{ri}} = \left( { - {{{f_k}} \mathord{\left/
 {\vphantom {{{f_k}} c}} \right.
 \kern-\nulldelimiterspace} c}} \right)\left( {{a_r} + {a_i}} \right)$,$\tilde a = as - {{\left( {{a_r} + {a_i}} \right)} \mathord{\left/
 {\vphantom {{\left( {{a_r} + {a_i}} \right)} 2}} \right.
 \kern-\nulldelimiterspace} 2}$, and discuss the transform results in different cases.

 \textbf{Case 1}: When ${a_r} = {a_i}$, ~\eqref{eq12} can be rewritten as
\begin{align}\label{eq16}
&Z\left[ {\left( {A_{J_{m,k}^{\left( r \right)}J_{m,k}^{\left( i \right)}}^C\left( {\Delta t,\tau } \right) + A_{J_{m,k}^{\left( i \right)}J_{m,k}^{\left( r \right)}}^C\left( {\Delta t,\tau } \right)} \right)} \right]\\ \notag
&= \exp \left( { - j2\pi {f_k}\frac{2}{c}\left[ {\left( {{v_r} + {v_i}} \right)\left( {\tau  + {t_d}} \right){\rm{ + }}\frac{{\left( {{a_r} + {a_i}} \right)t}}{s}} \right]} \right)\\ \notag
 &\times  2\alpha _{m,k}^{\left( r \right)}\alpha _{m,k}^{\left( i \right)} \times \cos \left[ {4\pi \Delta {v_{ri}}\frac{t}{{s\left( {\tau  + {t_d}} \right)}}} \right].
\end{align}

On this basis, the 2D-FFT is performed on the second row of~\eqref{eq16}, we can obtain
\begin{align}\label{eq17}
	\mathcal{F}_\tau  \{ \mathcal{F}_t &\left( \exp \left[- j2\pi f_k\frac{2}{c}(v_r + v_i)(\tau  + t_d) \right]  \right. \\ \notag
	&\times \left. \exp \left[- j2\pi f_k\frac{2}{c}(a_r + a_i)\frac{t}{s} \right] \right) \} \\
	\notag
	&= \exp \left(- j4\pi t_d f_k\frac{2v}{c} \right) \delta \left( - f_k\frac{2a}{c} - \frac{\nabla a_{ri}}{s}\right)\\
	\notag
	&\times \delta ( - f_k\frac{2v}{c} - \nabla v_{ri}).
\end{align}
After that, 2D-FFT is conducted on the third row of~\eqref{eq16} to get
\begin{align}\label{eq18}
	&\mathcal{F}_\tau \left\{ \mathcal{F}_t \left( 2\alpha _{m,k}^{(r)}\alpha _{m,k}^{(i)}\cos \left[ 4\pi \Delta v_{ri}\frac{t}{s(\tau  + t_d)} \right] \right) \right\}\\
	\notag
	&= 2\alpha _{m,k}^{(r)}\alpha _{m,k}^{(i)} \\
	\notag
	&\times \mathcal{F}_\tau \left\{ \delta \left[ -f_k\frac{a}{c} + \frac{\Delta v_{ri}}{s(\tau  + t_d)} \right] + \delta \left[ -f_k\frac{a}{c} - \frac{\Delta v_{ri}}{s(\tau  + t_d)} \right] \right\} \\ \notag \\
	&= \begin{cases}
		2\alpha _{m,k}^{(r)}\alpha _{m,k}^{(i)}  \cos \left( 4\pi \frac{\Delta v_{ri}}{sa}v \right) \\ \notag
        \\
		\times \frac{4\alpha _{m,k}^{(r)}\alpha _{m,k}^{(i)}\Delta v_{ri}}{sf_k^2\left( 4a^2/c^2 \right)}\exp \left( -j4\pi t_d f_k\frac{v}{c} \right), & \Delta v_{ri} \ne 0 \\ \notag
\\
		2\alpha _{m,k}^{(r)}\alpha _{m,k}^{(i)}\delta (v)\delta (a), & \Delta v_{ri} = 0.
	\end{cases}
\end{align}
On this basis, the two-dimensional convolution is conducted between~\eqref{eq17} and~\eqref{eq18} to obtain the transformation result of the cross term
\begin{align}\label{eq19}
	{G_{J_{m,k}^{(r)}J_{m,k}^{(i)}}}(v,a) &= 2\alpha _{m,k}^{(r)}\alpha _{m,k}^{(i)}\\
	\notag
	&\left\{ \begin{array}{l}
		\times 4s\alpha _{m1,k}^{(r)}\alpha _{m1,k}^{(i)}\exp \left( - j2\pi d{f_k}\frac{2v}{c} \right) \\
        \\
		\times\frac{\Delta {v_{lj}}}{{\tilde a}^2}\cos \left( 4\pi \frac{\Delta {v_{lj}}}{\tilde a}\tilde v \right),\quad {a_r} = {a_i} \ne as\\
        \\
		\times 0,\qquad \qquad \qquad \qquad \quad {a_r} = {a_i} = as.
	\end{array} \right.
\end{align}
\textbf{Case 2}: The transformation of \(A_{J_{m,k}^{(r)}J_{m,k}^{(i)}}^{C2}(\Delta t,\tau)\) with respect to \(\Delta t\) and \(\tau\) is an oscillatory integral, and the amplitude of the integrand exhibits slow variations in comparison to the oscillations controlled by the phase term. Hence, when \(a_r \ne a_i\), according to the stationary phase principle in~\cite{wong2001asymptotic}, the stationary point \(t'\) is determined as:
\begin{align}\label{eq20}
\frac{d}{{dt}}\left( {phase\left[ {A_{J_{m,k}^{\left( r \right)}J_{m,k}^{\left( i \right)}}^{C2}\left( {\Delta t,\tau } \right)} \right] + 4\pi \frac{{{f_k}a}}{c}t} \right) = 0.
\end{align}
On this basis, we can derive
\begin{align}\label{eq21}
t' = \left( {\frac{{ - 2{f_k}a}}{c} - \frac{{2\Delta {v_{ri}}}}{{s\tau '}}} \right)\frac{{{{\left( {s\tau '} \right)}^2}}}{{2\Delta {a_{ri}}}}.
\end{align}
Further, the second-order differential can be obtained
\begin{align}\label{eq22}
\frac{{{d^2}}}{{d{t^2}}}\left( {phase\left[ {A_{J_{m,k}^{\left( r \right)}J_{m,k}^{\left( i \right)}}^{C2}\left( {\Delta t,\tau } \right)} \right] + 4\pi \frac{{{f_k}a}}{c}t'} \right) = \frac{{4\pi \Delta {a_{ri}}}}{{ {{{\left( {s\tau '} \right)}^2}} }}.
\end{align}
Using~\eqref{eq21} and~\eqref{eq22} , we can obtain the scaled ambiguity function as follows
\begin{align}\label{eq23}
{\rm{SA}}{{\rm{F}}_{J_{m,k}^{\left( r \right)}J_{m,k}^{\left( i \right)}}}&\left( {\Delta t,\tau } \right) = {{{\mathcal F}}_t}\left( {A_{J_{m,k}^{\left( r \right)}J_{m,k}^{\left( i \right)}}^{C2}\left( {\Delta t,\tau } \right)} \right)\\ \notag
 &= \frac{{\left| {s\left( {\tau  + {t_d}} \right)} \right|}}{{\sqrt {2\left| {\Delta {a_{ri}}} \right|} }}\exp \left( {j{\mathop{\rm sgn}} \left( {\Delta {a_{ri}}} \right)\frac{\pi }{4}} \right)\\ \notag
 &\times \exp \left( {j\pi \left[ { - 4\Delta {v_{ri}}\frac{{s\left( {\tau  + {t_d}} \right)}}{{\Delta {a_{ri}}}}\frac{{{f_k}a}}{c}} \right]} \right)\\ \notag
 &\times \exp \left( {j\pi \left[ { - \frac{{{s^2}{{\left( {\tau  + {t_d}} \right)}^2}}}{{2\Delta {a_{ri}}}}\frac{{4f_k^2{a^2}}}{c}} \right]} \right)\\ \notag
 &\times \exp \left( {j\pi \left[ { - \frac{{2\Delta v_{ri}^2}}{{\Delta {a_{ri}}}} + \Delta {a_{ri}}\frac{{{{\left( {\tau  + {t_d}} \right)}^2}}}{2}} \right]} \right).
\end{align}
where ${\mathop{\rm sgn}} \left(  \cdot  \right)$ is the sign function.

\textbf{Case 2-1}: Under the condition of satisfying constraint \(a_r \ne a_i\), and further, if \(a = \pm \left| \Delta a_{ri}/{s} \right|\), then we have:
%\textbf{Case 2-1}: Under the condition of satisfying constraint ${a_r} \ne {a_i}$, and further, if $a =  \pm \left| {{{\Delta {a_{ri}}} \mathord{\left/
%			{\vphantom {{\Delta {a_{ri}}} s}} \right.
%			\kern-\nulldelimiterspace} s}} \right|$, then we have
\begin{align}\label{eq24}
  {{G'}_{J_{m,k}^{\left( r \right)}J_{m,k}^{\left( i \right)}}}\left( {v,a} \right) &= {{\mathcal{F}}_\tau }\left[ {{\text{SA}}{{\text{F}}_{J_{m,k}^{\left( r \right)}J_{m,k}^{\left( i \right)}}}\left( {\Delta t,\tau } \right)} \right] \hfill \notag \\
  &= -\frac{s}{{\sqrt {\left| {\Delta {a_{ri}}} \right|} }}\exp \left( { - j2\pi \frac{{\Delta v_{ri}^2}}{{\Delta {a_{ri}}}}} \right)  \notag \\
  &\times \exp \left[ {j\operatorname{sgn} \left( {\Delta {a_{ri}}} \right)\frac{\pi }{4}} \right]\\ \notag
   &\times\frac{c}{{4\sqrt 2 {\pi ^2}{f_k}\left( {{{sa\Delta {v_{ri}}} \mathord{\left/
 {\vphantom {{sa\Delta {v_{ri}}} {\Delta {a_{ri}}}}} \right.
 \kern-\nulldelimiterspace} {\Delta {a_{ri}}}} - v} \right)}} \hfill.
\end{align}
Leveraging the conjugation property of Fourier transform, we have
\begin{align}\label{eq25}
{{\mathcal{F}}_\tau }\left[ {{{\mathcal{F}}_t}\left( {A_{J_{m,k}^{\left( r \right)}J_{m,k}^{\left( i \right)}}^{C2 * }\left( {\Delta t,\tau } \right)} \right)} \right] = {G'^*}_{J_{m,k}^{\left( r \right)}J_{m,k}^{\left( i \right)}} \left( { - v, - a} \right),
\end{align}
where ${\left( {} \right)^ * }$ is the complex conjugate operator. Therefore, using~\eqref{eq25} and modulation property, we have
\begin{align}\label{eq26}
{{\mathcal{F}}_\tau }&\left\{ {{{\mathcal{F}}_t}\left[ {2A_{J_{m,k}^{\left( r \right)}J_{m,k}^{\left( i \right)}}^{C1}\left( {\Delta t,\tau } \right) \times \operatorname{Re} \left\{ {A_{J_{m,k}^{\left( r \right)}J_{m,k}^{\left( i \right)}}^{C2}\left( {\Delta t,\tau } \right)} \right\}} \right]} \right\} \notag \\
&= {G'_{J_{m,k}^{\left( r \right)}J_{m,k}^{\left( i \right)}}}\left( {\frac{{ - 2{f_k}v}}{c} - \nabla {v_{ri}},\frac{{ - 2{f_k}a}}{c} - \frac{{\nabla {a_{ri}}}}{s}} \right) \\ \notag
&+ {G'^*}_{J_{m,k}^{\left( r \right)}J_{m,k}^{\left( i \right)}} \left( {\frac{{2{f_k}v}}{c} + \nabla {v_{ri}},\frac{{2{f_k}a}}{c} + \frac{{\nabla {a_{ri}}}}{s}} \right) \hfill.
\end{align}

At last, we substitute~\eqref{eq24} into~\eqref{eq26} and utilize the translation and scaling properties to~\eqref{eq26}, we can obtain
 \begin{align}\label{e27}
{G_{J_{m,k}^{\left( r \right)}J_{m,k}^{\left( i \right)}}}\left( {v,a} \right)&{\rm{ = }} - \frac{{s\alpha _{m1,k}^{\left( r \right)}\alpha _{m1,k}^{\left( i \right)}}}{{{{\left| {\Delta {a_{ri}}} \right|}^{{1 \mathord{\left/
 {\vphantom {1 2}} \right.
 \kern-\nulldelimiterspace} 2}}}}} \\ \notag
 &\times \cos \left( {\left[ {{\mathop{\rm sgn}} \left( {\Delta {a_{ri}}} \right)\frac{\pi }{4} - 2\pi \frac{{\Delta v_{ri}^2}}{{\Delta {a_{ri}}}}} \right]} \right) \\ \notag
 &\times \frac{{\exp \left( {j2\pi {t_d}\left( {{{ - {f_k}2v} \mathord{\left/
 {\vphantom {{ - {f_k}2v} c}} \right.
 \kern-\nulldelimiterspace} c}} \right)} \right)}}{{\sqrt 2 {\pi ^2}{{\left( {\tilde v - \frac{{\Delta {v_{ri}}}}{{\Delta {a_{ri}}}}\left( { \pm \left| {\Delta {v_{ri}}} \right| - \nabla {v_{ri}}} \right)} \right)}^2}}}.
\end{align}

\begin{figure}[tbp!]
  \centering
  \includegraphics[height=2.5cm]{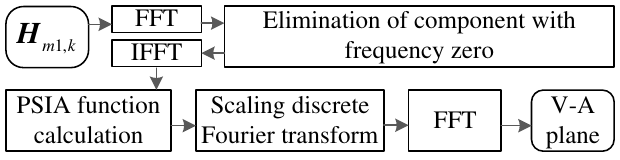}
  \caption{Flowchart of the proposed velocity and acceleration estimation algorithm.}
  \label{AFLC}
\end{figure}

\textbf{Case2-2}: Through the same way, when ${a_r} \ne {a_i}$, $a \ne  \pm {{\left| {\Delta {a_{ri}}} \right|} \mathord{\left/ {\vphantom {{\left| {\Delta {a_{ri}}} \right|} s}} \right.  \kern-\nulldelimiterspace} s}$, and $\tilde a \ne  \pm \left| {\Delta {a_{ri}}} \right|$, we have
\begin{align}\label{eq28}
{G_{J_{m,k}^{\left( r \right)}J_{m,k}^{\left( i \right)}}}\left( {v,a} \right) &= \frac{{\left| {\tilde v\Delta {a_{ri}} - \tilde a\Delta {v_{ri}}} \right|}}{{{{\left| {{{\tilde a}^2} - \Delta a_{ri}^2} \right|}^{{3 \mathord{\left/
 {\vphantom {3 2}} \right.
 \kern-\nulldelimiterspace} 2}}}}}   \\ \notag
& \times 4\alpha _{m,k}^{\left( r \right)}\alpha _{m,k}^{\left( i \right)}\exp \left( {j2\pi {t_{d}}\frac{{ - {f_k}2v}}{c}} \right) \\ \notag
 & \times \cos \left\{ {\frac{\pi }{4}{\rm{sgn}}\left( {\Delta {a_{ri}}} \right)  \left[ {1 + {\rm{sgn}}\left( {\Delta a_{ri}^2 - {{\tilde a}^2}} \right)} \right]} \right.\\ \notag
&\left. { - \frac{{2\pi \Delta {a_{ri}}}}{{\Delta a_{ri}^2 - {{\tilde a}^2}}}{{\left( {\tilde v - \frac{{\tilde a\Delta {v_{ri}}}}{{\Delta {a_{ri}}}}} \right)}^2} - 2\pi \frac{{\Delta v_{ri}^2}}{{\Delta {a_{ri}}}}} \right\}.
\end{align}
For the remaining cases where ${a_r} \ne {a_i}$, $a \ne  \pm {{\left| {\Delta {a_{ri}}} \right|} \mathord{\left/ {\vphantom {{\left| {\Delta {a_{ri}}} \right|} s}} \right. \kern-\nulldelimiterspace} s}$, and ${\rm{ }}\tilde a =  \pm \left| {\Delta {a_{ri}}} \right|$, we have ${G_{J_{m,k}^{\left( r \right)}J_{m,k}^{\left( i \right)}}}\left( {v,a} \right){\rm{ = 0}}$. In Fig.~\ref{AFLC}, we briefly summarize the process of generating the V-A plane, i.e., the estimation process of velocity and acceleration.

From the derivation process presented above, we can obtain several key findings.
\begin{itemize}
\item [1)]
The auto terms presented in \eqref{eq13}, which correspond to the DPs, appear as impulses on the generated V-A plane, and the corresponding coordinate values are the estimated velocity and acceleration.
\item [2)]
Between two auto terms, the cross term appears as a cosine structure, which oscillates perpendicular to the line connecting two auto terms, with the envelope determined by the auto terms and their positions.
\item [3)]
The strength of the cross term is smaller than that of the auto term. Therefore the auto term can be identified from the V-A plane.
\end{itemize}

\begin{figure}[tbp!]
  \centering
  \includegraphics[height=5cm]{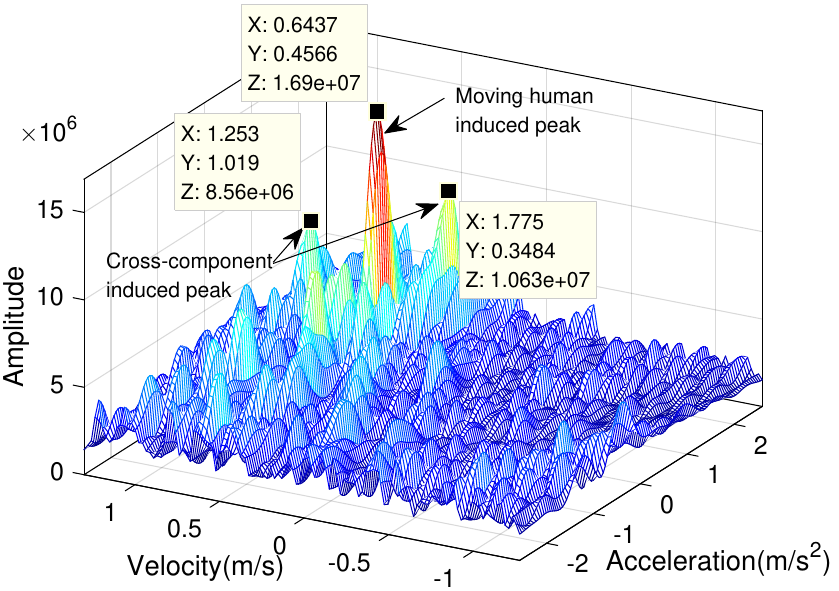}
  \caption{An example of velocity and acceleration estimation result. From the V-A plane, we can see that the direct path and static reflection paths, with both velocity and acceleration equal to zero, are filtered out, leaving only the DP in the V-A plane. Besides, we can also observe that the energy of the auto term is stronger than that of cross terms (caused by the parametric symmetric instantaneous auto-correlation function calculation) and cross component (introduced by conjugate multiplication). This leads to an easy identification of the peak corresponding to DP, thereby facilitating the estimation process.}
  %\vspace{-0.5cm}
  \label{Ac}
\end{figure}

However, ${{\bf{J}}_{m,k}}$ includes not merely the DPs, but also the cross component caused by the conjugate multiplication. This indicates that the V-A plane contains the interference terms introduced by the cross component, as expressed in~\eqref{eq3}. Fortunately, the amplitude of the cross component is smaller than that of the DP. Hence, the auto term of the DP can still be distinguished by analyzing the amplitude. Figure~\ref{Ac} gives an example of velocity and acceleration estimation result, i.e., the V-A plane, based on CSI collected from a real-world scenario where one moving human appears in the monitored area. It is clear that there are multiple interference terms, which manifest as raised peaks, in the generated V-A plane. However, their magnitudes are lower than that of the DP induced by the moving human, making the DP easy to be recognized.

While effective, the performance of DP-AcE may degrade when multiple individuals appear, as the auto term produced by one DP may be weaker than the interference caused by another DP. In this case, we can use a step-wise estimation-elimination strategy~\cite{wang2019csi} to solve the problem. For instance, when two moving humans appear, DP-AcE first estimates the velocity and acceleration of the one generating the strongest peak in the V-A plane. Then, using the estimation results and the signal model, the CSI corresponding to this DP is reconstructed and removed from received CSI, making the DP caused by the other moving human the dominant one in the residual CSI. Hence, the velocity and acceleration of the other moving human can be estimated.

\subsection{Fall Detection }
Based on estimated acceleration, this paper illustrates the significance of acceleration estimation in wireless sensing by using fall detection as an example. Utilizing the ${\bf{J}}_{m,k}$ obtained from the data in the window, the velocity and acceleration of the DPLC can be estimated using the algorithm described in Section B. By sliding the window, continuous estimation can be achieved. Nevertheless, noise in the data prevents the original estimation results from being directly used for fall detection. Therefore, prior to that, the estimation results are filtered and segmented. It is worth noting that while velocity and acceleration both describe human movement, the acceleration has been shown to be better than velocity in detecting fall~\cite{hu2020wifi}. For this reason, only estimated acceleration is utilized for fall detection in this paper.
\subsubsection{Data Filtering and Segmentation}
\begin{figure}
	\centering
	\subfigure[The original acceleration trace.]
    { \begin{minipage}{8.5cm}
    \includegraphics[width=\textwidth]{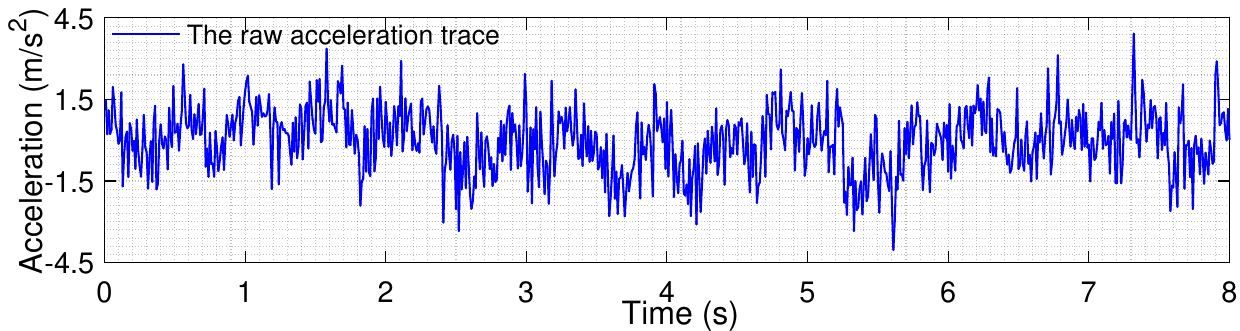}
		\end{minipage}
	}
	\subfigure[The filtered and segmented acceleration trace.]
 {
		\begin{minipage}{8.5cm}
		\includegraphics[width=\textwidth]{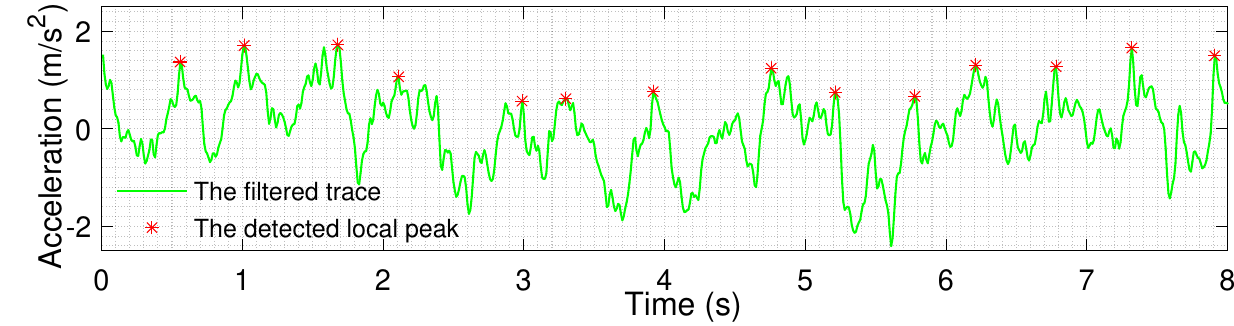}
		\end{minipage}
	}
\caption{The original acceleration estimation results and the processed estimation results.}
\label{filtered}
%\vspace{-0.5cm}
\end{figure}

Given that the frequency range of signal fluctuations, which are caused by human actions in indoor environments, is limited, DP-AcE first uses a low-pass filter to filter out the out-of-band noise. After that, the $\ell_1$ trend filter~\cite{kim2009ell_1} is used to smooth the acceleration trace. Assuming the acceleration trace with $U$ samples after the low-pass filtering is ${\bf{a}} = [{a_1}, \ldots,  {a_u}, \ldots, {a_U}]$, then the smoothing process is described as
\begin{align}
\mathop {\min }\limits_{{a'_u},\forall u} \sum\limits_{u = 1}^U {{{\left( {{a'_u} - {a_u}} \right)}^2}}  + \xi \sum\limits_{u = 2}^{U - 1} {\left| {{a'_u-1} - 2{a'_u} + {a'_u+1}} \right|},
\end{align}
where ${a'_u}$ is the smoothed acceleration, $\xi $ is the regularization parameter utilized to balance the trade-off between the smoothness and $\left| {{a'_u} - {{a}_u}} \right|$. Figure~\ref{filtered} gives an example of the original estimated acceleration trace and the processed trace. Comparing the original trace in Fig.~\ref{filtered}(a) with the processed trace in Fig.~\ref{filtered}(b), one can see that the processed data contains less noise, resulting in a clear pattern of acceleration changes triggered by human walking.

Following the filtering, data segmentation is carried out through the extraction of local peaks from the filtered acceleration trace. Here, automatic multiscale-based peak detection algorithm~\cite{scholkmann2012efficient} is adopted to achieve this goal, which works as follows. First, the local maxima ${b_{q,u}}$ of the smoothed trace ${\bf{ a'}} = [{ a'_1}, \cdots { a'_u} \cdots { a'_U}]$ signal is determined using a sliding window with varying length. On this basis, the local maxima scalogram can be constructed as
\begin{align}
{\bf{B}} = \left[ {\begin{array}{*{20}{c}}
{{b_{1,1}}}& \cdots &{{b_{1,U}}}\\
 \vdots & \ddots & \vdots \\
{{b_{Q,1}}}& \cdots &{{b_{Q,U}}}
\end{array}} \right].
\end{align}
Then, a row-wise summation is performed on ${\bf{B}}$ to get $\left[ {{\eta _1},{\eta _1} \cdots {\eta _Q}} \right]$, where $Q = \left\lceil {{U \mathord{\left/
 {\vphantom {U 2}} \right.
 \kern-\nulldelimiterspace} 2}} \right\rceil  - 1$ and $\left\lceil  \cdot  \right\rceil $ is the ceiling operator. After that, the global minimum of $\left[ {{\eta _1},{\eta _2} \cdots {\eta _Q}} \right]$, denoted as ${\eta _{\min }}$, can be extracted. Through this parameter, ${\bf{B}}$ is reshaped to ${\bf{B'}}$, by removing the ${b_{q,u}}$ that $q > {\eta _{\min }}$.
At last, the peaks are detected by calculating the column-wise standard deviation of ${\bf{B'}}$ via
\begin{align}
{\sigma _u} = \frac{1}{{\eta  - 1}}{\left[ {\sum\limits_{q = 1}^\eta  {{{\left( {{{b'}_{q,u}} - \frac{1}{\eta }\sum\limits_{q = 1}^\eta  {{{b'}_{q,u}}} } \right)}^2}} } \right]^{{1 \mathord{\left/
 {\vphantom {1 2}} \right.
 \kern-\nulldelimiterspace} 2}}},
\end{align}
and finding all indices $u$ for which ${\sigma _u} = 0$ holds. Taking the smoothed data as an example, Fig.~\ref{filtered}(b) shows the detected local peaks in red.

\subsubsection{Feature Extraction and Fall Detection}
\begin{figure}[tbp!]
  \centering
  \includegraphics[height=4.3cm]{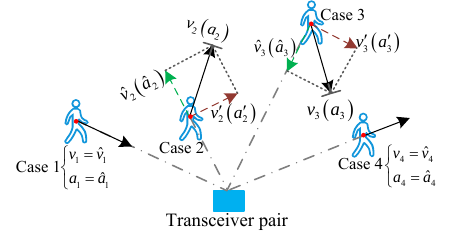}
  \caption{The relationship between the estimated velocity and acceleration and the true acceleration and velocity. Only in cases 1 and 4, the algorithm is able to accurately estimate the true acceleration (and velocity) of the moving human; in other cases, the estimated acceleration (and velocity) is only a component of the true acceleration (and velocity), as indicated by the green dashed line.}
  \label{F4}
  %\vspace{-0.5cm}
\end{figure}

Falls, such as slips or stumbles, are characterized by the abrupt transition of the human body from an upright standing position to a lying-down position, without any voluntary control. During this transition, the body experiences a sudden change in acceleration, marked by rapid acceleration and deceleration, followed by a complete halt in velocity and acceleration when the body hits the ground~\cite{zhang2018wispeed}. Based on this feature, fall detection can be achieved by comparing changes in acceleration to predetermined thresholds, provided that the absolute acceleration of the human can be accurately captured. However, the proposed method estimates the acceleration of DPLC, which equals to the absolute acceleration of the moving human target only when the target is moving towards the signal transceiver pair. This implies that if the human is not moving towards the transceiver pair, which usually happens, the estimated velocity and acceleration are the components of the true absolute acceleration and velocity. Figure~\ref{F4} illustrates the relationship between the estimation results and the true acceleration and velocity.

Taking the above discussion into consideration, we extract a series of relative statistical features between adjacent data segments from three perspectives and train an SVM-based classifier to realize fall detection. More concretely, the relative truncated mean (rTM), mode, and median, which describe the central tendency of the $w$-th ($w>1$) data segment with respect to its adjacent data segment, are extracted first. For instance, to calculate rTM, we utilize the data from which the samples with the lowest 5\% and highest 5\% values are discarded, and calculates the truncated mean. Assuming the truncated mean of $w$-th data segment is ${T_w}$, then the rTM of the $w$-th data segment can be calculated as
\begin{align}
rT{M_w}{\rm{ = }}\frac{{{{{T_w}} \mathord{\left/
 {\vphantom {{{T_w}} {{T_{w - 1}}}}} \right.
 \kern-\nulldelimiterspace} {{T_{w - 1}}}} + {{{T_w}} \mathord{\left/
 {\vphantom {{{T_w}} {{T_{w + 1}}}}} \right.
 \kern-\nulldelimiterspace} {{T_{w + 1}}}}}}{2}.
\end{align}
Similarly, the relative median and mode can be obtained in the same way by replacing ${T_w}$ with the median and mode value of the $w$-th data segment, respectively.

Besides the central tendency, the relative extreme deviation, standard deviation, and inter-quartile range (IQR), which depict the dispersion of the $w$-th data segment with respect to its adjacent data segment, are then extracted. Taking the relative IQR as an example,
its calculation process corresponding to $w$-th data segment is defined as
\begin{align}
rIQ{R_w}{\rm{ = }}{{\left( {\frac{{T{Q_w} - F{Q_w}}}{{T{Q_{w + 1}} - F{Q_{w + 1}}}} + \frac{{T{Q_w} - F{Q_w}}}{{T{Q_{w - 1}} - F{Q_{w - 1}}}}} \right)} \mathord{\left/
 {\vphantom {{\left( {\frac{{T{Q_w} - F{Q_w}}}{{T{Q_{w + 1}} - F{Q_{w + 1}}}} + \frac{{T{Q_w} - F{Q_w}}}{{T{Q_{w - 1}} - F{Q_{w - 1}}}}} \right)} 2}} \right.
 \kern-\nulldelimiterspace} 2},
\end{align}
where $F{Q_w}$ and $T{Q_w}$ represent the first and third quartile of $w$-th segment data, respectively.

In addition to the above-mentioned features, the skewness coefficient and kurtosis coefficient are also computed. For the smoothed acceleration, specifically, the skewness coefficient is a measure of distribution asymmetry, which is defined as
\begin{align}
Sk = \frac{{{{\sum\limits_{x = 1}^{\# \left\{ w \right\}} {{{\left( {{{a'}_x} - \bar a'} \right)}^3}} } \mathord{\left/
 {\vphantom {{\sum\limits_{x = 1}^{\# \left\{ w \right\}} {{{\left( {{{a'}_x} - \bar a'} \right)}^3}} } {\# \left\{ w \right\}}}} \right.
 \kern-\nulldelimiterspace} {\# \left\{ w \right\}}}}}{{{{\left[ {{{\sum\limits_{x = 1}^{\# \left\{ w \right\}} {{{\left( {{{a'}_x} - \bar a'} \right)}^2}} } \mathord{\left/
 {\vphantom {{\sum\limits_{x = 1}^{\# \left\{ w \right\}} {{{\left( {{{a'}_x} - \bar a'} \right)}^2}} } {\left( {\# \left\{ w \right\}{\rm{ - }}1} \right)}}} \right.
 \kern-\nulldelimiterspace} {\left( {\# \left\{ w \right\}{\rm{ - }}1} \right)}}} \right]}^{{3 \mathord{\left/
 {\vphantom {3 2}} \right.
 \kern-\nulldelimiterspace} 2}}}}} = \frac{{{{\hat \mu }_3}}}{{{{\hat \sigma }^3}}},
\end{align}
where $\# \left\{ w \right\}$ is the sample size of the $w$-th acceleration data segment, $\bar a'$ and $\hat \sigma $ are the sample mean value and standard deviation, ${\hat \mu _3}$ is the 3rd-order central moment. A positive $Sk$ indicates a longer tail to the right, while a negative one represents a longer tail to the left, with respect to the standard normal distribution. The kurtosis coefficient, which demonstrates whether the density of the sample is more or less peaked around its center than the density of a normal curve, is calculated as
\begin{align}
Kt = \frac{{{{\sum\limits_{x = 1}^{\# \left\{ w \right\}} {{{\left( {{{a'}_x} - \bar a'} \right)}^4}} } \mathord{\left/
 {\vphantom {{\sum\limits_{x = 1}^{\# \left\{ w \right\}} {{{\left( {{{a'}_x} - \bar a'} \right)}^4}} } {\# \left\{ w \right\}}}} \right.
 \kern-\nulldelimiterspace} {\# \left\{ w \right\}}}}}{{{{\left( {{{\sum\limits_{x = 1}^{\# \left\{ w \right\}} {{{\left( {{{a'}_x} - \bar a'} \right)}^2}} } \mathord{\left/
 {\vphantom {{\sum\limits_{x = 1}^{\# \left\{ w \right\}} {{{\left( {{{a'}_x} - \bar a'} \right)}^2}} } {\# \left\{ w \right\}{\rm{ - }}1}}} \right.
 \kern-\nulldelimiterspace} {\# \left\{ w \right\}{\rm{ - }}1}}} \right)}^2}}},
\end{align}
where ${\hat \mu _4}$ is the 4th-order central moment. A positive $Kt$ indicates that the sample density is sharper around its center than that of standard normal curve, while a negative one means flatter.

During normal walking, gait parameters such as walking velocity, acceleration, step length, and step frequency are typically stable, and their statistical characteristics are consistent across different data segments. However, these parameters exhibit varying degrees of fluctuation during falls, resulting in changes in their statistical properties. Hence, the aforementioned features serve as a comprehensive representation of the relative statistical characteristics among data segments that can be used to detect falls in dynamic scenarios. More specifically, the extracted features are compiled into a vector and fed to a binary SVM with a radial basis function kernel for training, using the LIBSVM tool~\cite{fan2005working}. During the training process, the SVM maps objective and non-objective data to a high-dimensional space via the kernel function and finds a hyperplane with the largest margin, which is used for classification during the detection process. Notably, the features employed in this study are derived solely from the acceleration of the moving human and are determined purely by the body's own state, independent of the surrounding environment. Therefore, the trained classifier holds the potential to be trained once and applied in multiple scenarios, thereby offering greater generalization than some other existing machine-learning-based fall detection approaches.
\subsection{Complexity Analysis}
In the acceleration estimation process, DP-AcE includes five main steps, including conjugate multiplication based phase error elimination, strong signal interference elimination, parametric symmetric instantaneous auto-correlation calculation, decoupling, and 2D-FFT. If the CSI data stream contains $W$ measurements and $L$ zeros are padded in the FFT process, the overall complexity of the acceleration estimation stage is $O\left( {{W^2}\left( {7 + 5{{\log }_2}W} \right) + WL\left( {1 + \frac{3}{2}{{\log }_2}L} \right)} \right)$. The fall detection process mainly involves the SVM training and decision-making. Let ${{N_S}}$ denote the number of support vectors, $ {{D_L}} $ is the dimension of the input vector, $ {{N_T}}$ is the number of training samples, and $ {{C_T}} $ is the number of operations in the SVM kernel, then the computational complexity of training and decision-making are $O\left( {N_S^3 + {N_S}{N_T} + {N_s}{D_L}{N_T}} \right)$ and $O\left( {{C_T}{N_S}} \right)$, respectively~\cite{jakkula2006tutorial}.

\section{IMPLEMENTATION AND EVALUATION}
In this section, we prototype the proposed system on commodity WiFi devices and conduct comprehensive experiments to evaluate its performance under different real-world scenarios. We first introduce the experimental environment, hardware configuration, experimental methods, and evaluation metrics. Subsequently, we evaluate the system from two perspectives: parameter estimation accuracy and fall detection performance.
\begin{figure*}[htbp]
%\vspace{-.1cm}
\centering
%\hspace{-.5cm}
\subfigure[The conference room.]{
\begin{minipage}[t]{0.3\linewidth}
\centering
\includegraphics[width=5.5cm]{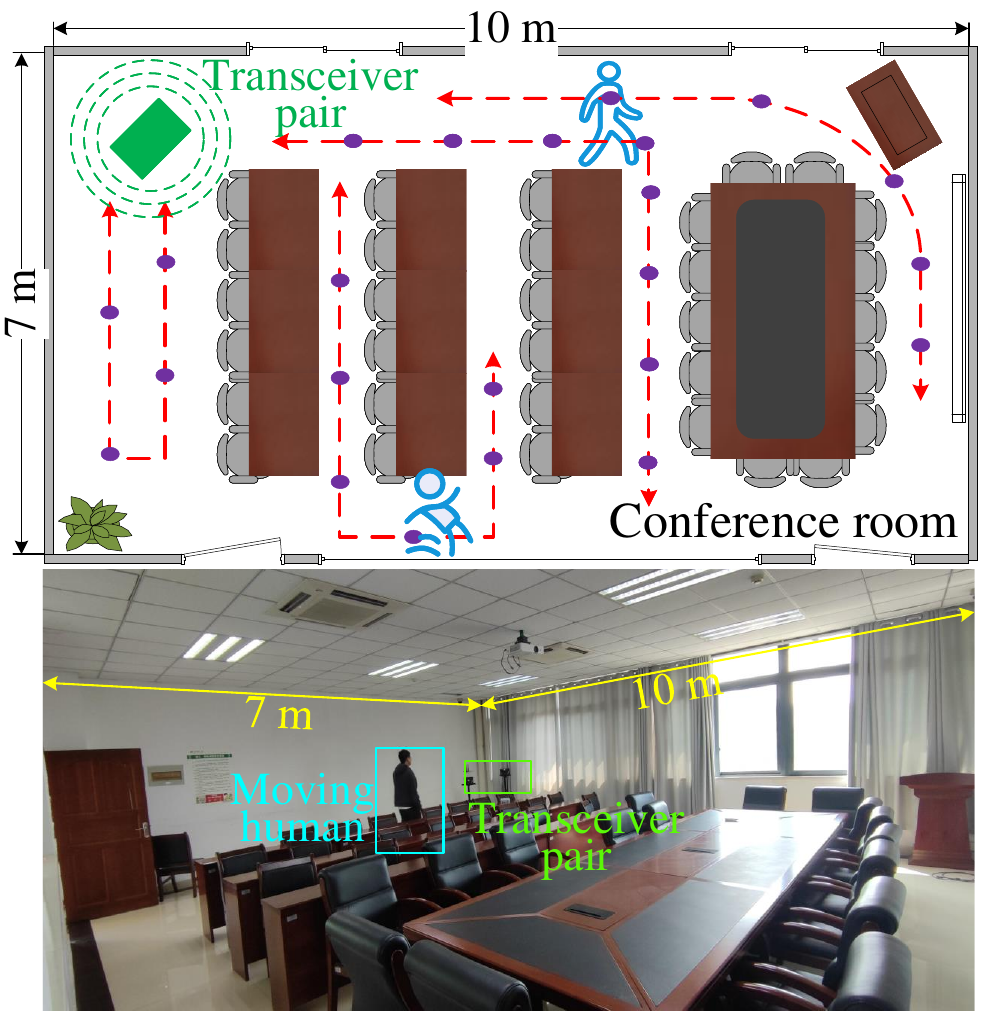}
%\caption{fig1}
\end{minipage}%
}%
\subfigure[The corridor.]{
\begin{minipage}[t]{0.3\linewidth}
\centering
\includegraphics[width=5.8cm]{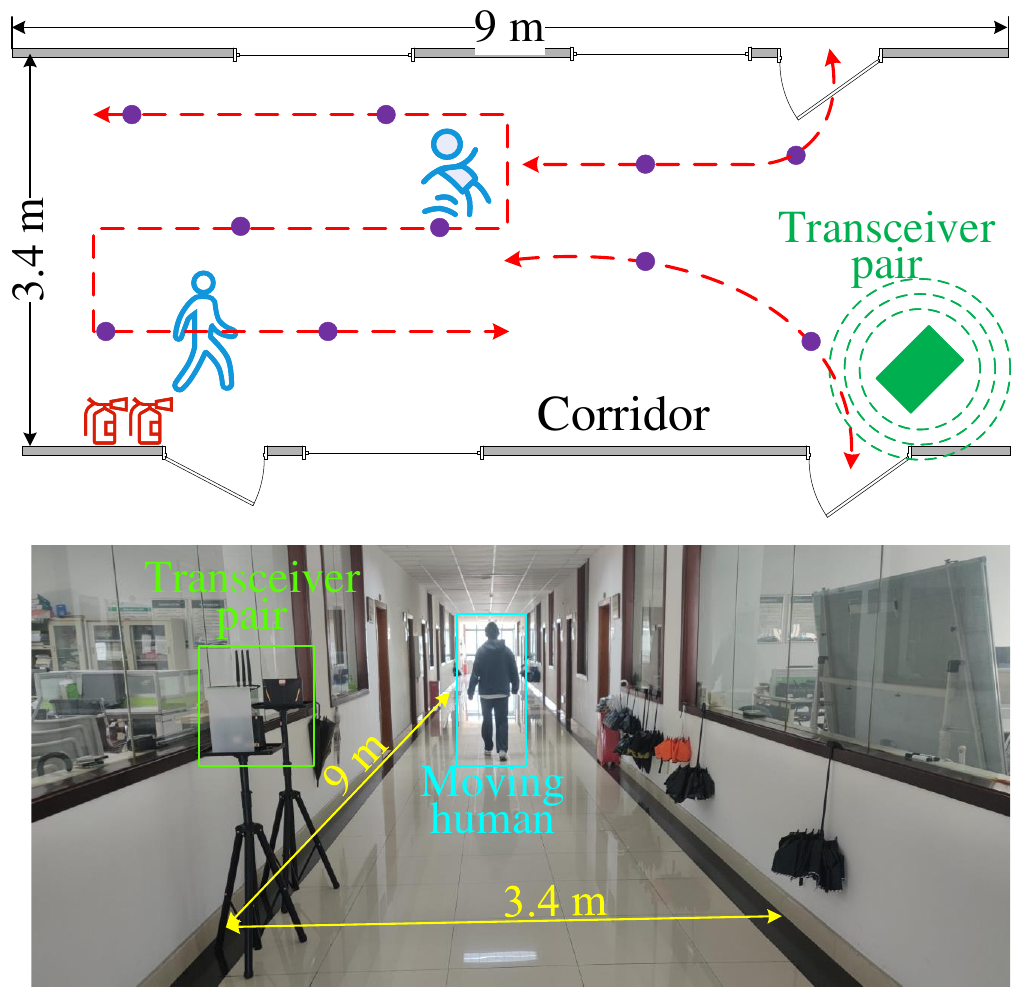}
%\caption{fig2}
\end{minipage}%
}%
\subfigure[The experimental hardware.]{
\begin{minipage}[t]{0.4\linewidth}
\centering
\includegraphics[width=6.2cm]{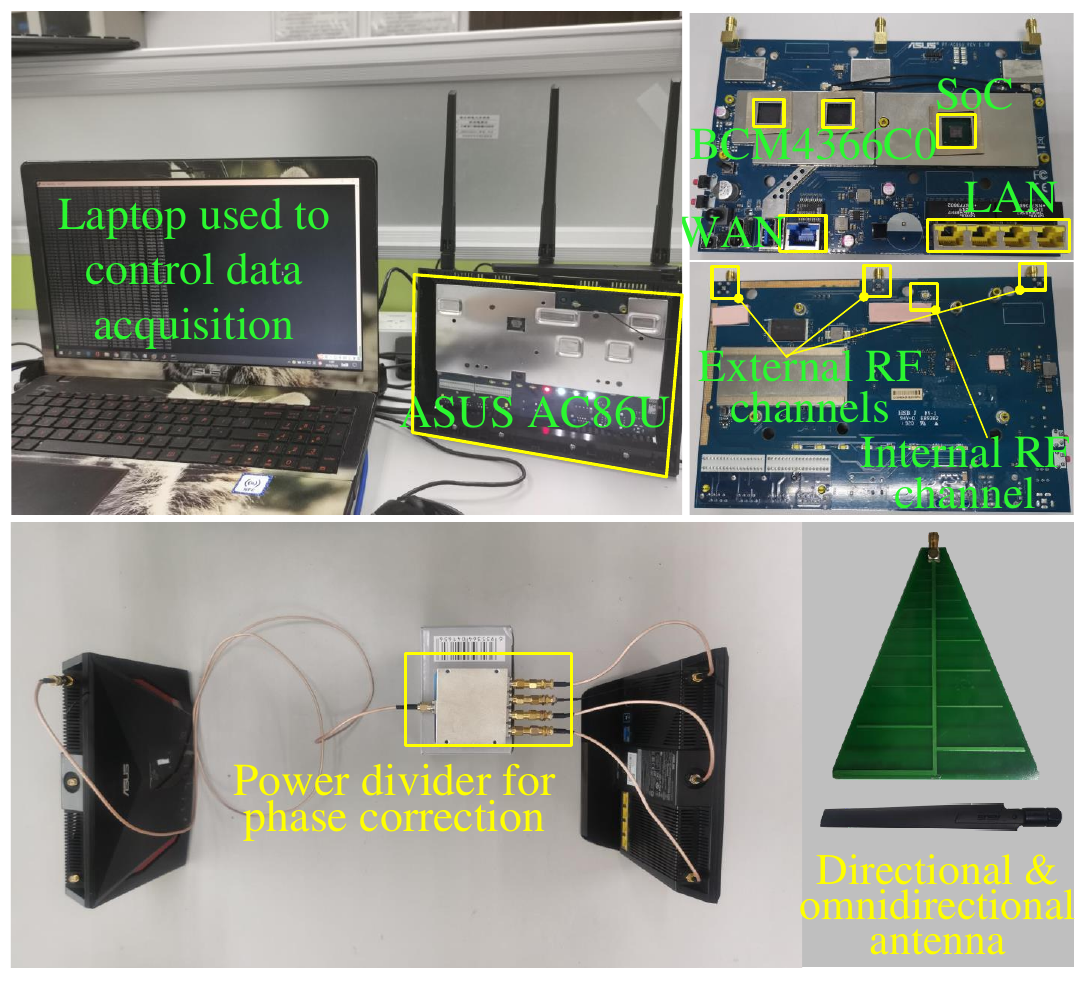}
%\caption{fig2}
\end{minipage}
}%
\centering
\caption{The test scenarios and hardware configuration. In the test scenarios, the red dashed line represents the preset path of human movement, and the purple dots represent the preset locations where fall happens. In the hardware diagram in (c), we label the chip, radio frequency (RF) channels, and other components. During the data collection process, a laptop is used to send instructions to the device to control the signal parameters for CSI collection.}
\label{Hardware}
%\vspace{-0.5cm}
\end{figure*}
\subsection{Implementation}
\subsubsection{ Experimental Scenarios} To verify the effectiveness of the proposed system, we conduct experiments in two typical indoor scenarios. The first scenario is a 10 m $\times$ 7 m conference room, which contains rows of tables, chairs, and other furniture made of wood and metal. The second scenario is a 9 m $\times$ 3.4 m corridor. In both experimental scenarios, a large number of multipath signals exist due to the presence of static objects such as walls and furniture. In comparison, the corridor appears relatively emptier than the conference room, as there are a few static objects that can produce strong reflections other than the walls. The specific floorplan is presented in Figs.~\ref{Hardware} (a) and (b).

\subsubsection{Hardware Configurations} For the experiments, two access points equipped with Broadcom 4366C0 chips and the Nexmon toolkit~\cite{schulz2016using} are used to construct the transmission link and collect CSI under the IEEE 802.11ac protocol. The center frequency of the signal is 5.805 GHz, the signal bandwidth is 80 MHz (including 256 subcarriers), and the default transmission rate is set at 600 packets/s. The Tx and Rx are placed together, about 0.4 m apart and 1.2 m above the ground. The Tx uses one antenna to send the signal, while the Rx utilizes four antennas\footnote{The access point adopted here contains four radio frequency channels, three of which are external and one is internal.} to collect data. One directional antenna is used to receive the reference signal, which travels from the Tx to the Rx directly, and the other antennas are omnidirectional and used for receiving the surveillance signal. The specific devices used in the experiments are shown in Fig.~\ref{Hardware}. Besides that, the initial phase differences among different radio frequency channels are measured and compensated with the help of a power divider before the experiment. The collected experimental data is processed by the server via MATLAB in a non-real-time manner, following the sequence of acceleration estimation, feature extraction, and fall detection. Unless specifically mentioned, the default setup introduced for the performance evaluation is used throughout the rest of this paper. Based on the above configuration, two points need to be noted.
\begin{itemize}
\item The space around the transmitting antenna can be divided into three regions, i.e., reactive near-field region, Fresnel region, and the Fraunhofer region, arranged from near to far based on the distance from the antenna. The reactive near-field region is defined as that region of the field immediately surrounding the antenna wherein the reactive field predominates~\cite{balanis1992antenna}. According to the definitions in~\cite{Standard}, the outer boundary of the near-field is taken to exist at a distance of $ 0.62\sqrt {{{D_a^3} \mathord{\left/
 {\vphantom {{D_a^3} \lambda }} \right.
 \kern-\nulldelimiterspace} \lambda }}$ from the antenna, where $\lambda$ is the wavelength and ${D_a}$ is the maximum dimension of the antenna. In this paper, the signal wavelength is approximately 0.051 m and the distance between the Tx and Rx is about 0.4 m. Since a single antenna is used for signal transmission, there are no near-field effects.
\item According to the `Doppler Power Spectrum and Channel Coherence Time' in~\cite{goldsmith2005wireless}, channel parameters remain stable within the channel coherence time. Based on existing studies, the channel coherence time for indoor scenarios is around several hundred milliseconds. This conclusion has been applied to related studies, including indoor localization~\cite{vasisht2016decimeter}, where the channel coherence time is approximately 250 ms. In this paper, the transmission rate is 600 packets/s, and the length of the sliding window is 120, which occupies a duration of 200 ms. Therefore, the channel parameters are considered to be stable during the estimation process.
\end{itemize}

\subsubsection{Experimental Method} To evaluate the parameter estimation accuracy, the tester is asked to walk along the predefined tracks to collect the data, based on which the velocity and acceleration can be estimated. Then, the estimation results are converted into distances. Therefore, we calculate the error between the computed walking distance and the absolute distance of preset tracks to evaluate the parameter estimation performance of the proposed algorithm.

To evaluate fall detection performance, data is first collected in the conference room for classifier training. The training set includes data of walking, falling, sitting down, and standing up. During training, falling data is marked as a positive category, while the rest is noted as a negative category. Based on the trained classifier, the fall detection performance is then evaluated in both the conference room and corridor. In the single-target cases, the tester continuously performs a series of actions, including walking, falling, standing up, and sitting down, in both test scenarios, to complete the data collection. In the multi-target cases, one person is asked to perform continuous actions, while the others perform continuous actions except falling, to complete the data collection. Based on the collected test data, the system's fall detection performance is evaluated.

\subsubsection{Evaluation Metrics} The performance of the DP-AcE based fall detection is measured by the true positive rate (TPR) and the false positive rate (FPR). The TPR refers to the probability that the fall is correctly detected, i.e., a fall happens and the system reports the occurrence of the fall. The FPR indicates the probability that the fall does not happen, yet the system reports it. It is worth noting that for the proposed system, a correct detection in the case of multiple targets means successfully detecting the fall and identifying which individual has fallen.

In terms of acceleration estimation, we compare DP-AcE with Wispeed~\cite{zhang2018wispeed} and the MUSIC based method~\cite{li2017indotrack}. Wispeed establishes a relationship between the walking velocity and CSI autocorrelation function to estimate velocity and then computes the acceleration by differentiating the estimated velocity. The other method employs the MUSIC algorithm to estimate the walking velocity, and then we differentiate the estimated speed to obtain the acceleration. For fall detection performance, we compare our method with RT-Fall~\cite{wang2016rt}, FallDeFi~\cite{palipana2018falldefi}, and Wispeed~\cite{zhang2018wispeed}. RT-Fall segments the CSI data based on phase differences and extracts eight features to realize the fall detection via a trained SVM classifier. FallDeFi extracts features extracted from the STFT spectrum and power burst curve, and utilizes a sequential forward selection algorithm to single out features that are resilient to changes in the environment. Then, these features are used to train an SVM classifier for fall detection. Alternatively, Wispeed compares the estimated velocity and acceleration with pre-set thresholds to perform fall detection.

\subsection{Parameter Estimation Performance Evaluation}
\begin{figure*}[htbp]
%\vspace{-.1cm}
%\vspace{-0.5cm}
\centering
%\hspace{-.5cm}
\subfigure[ ]{
\begin{minipage}[t]{0.3\linewidth}
\centering
\includegraphics[width=4.8cm]{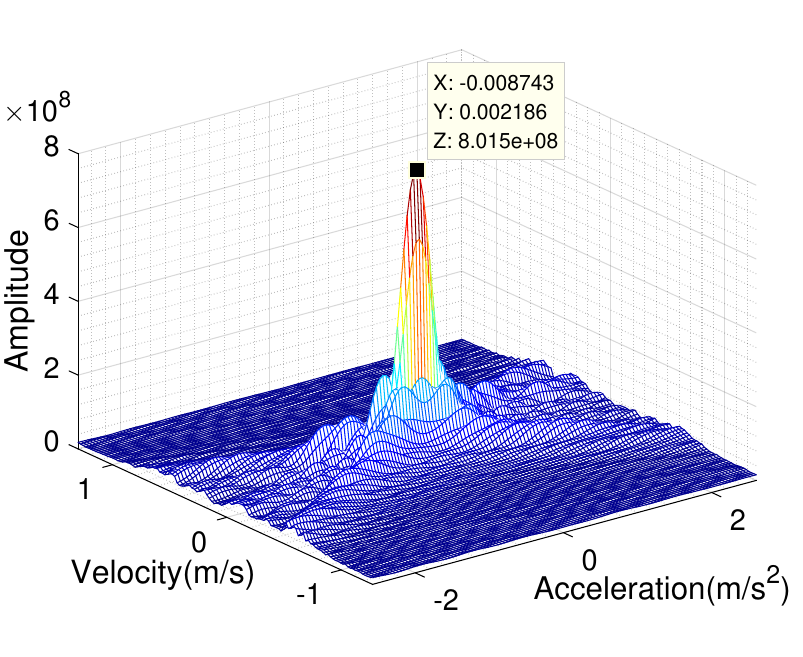}
%\caption{fig1}
\end{minipage}%
}%
\subfigure[ ]{
\begin{minipage}[t]{0.35\linewidth}
\centering
\includegraphics[width=4.8cm]{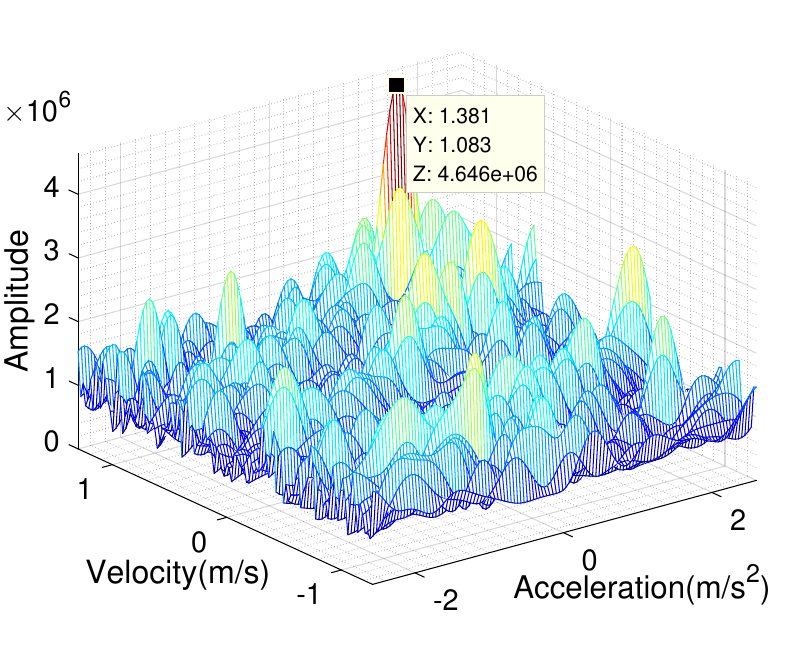}
%\caption{fig2}
\end{minipage}%
}%
\subfigure[ ]{
\begin{minipage}[t]{0.3\linewidth}
\centering
\includegraphics[width=4.8cm]{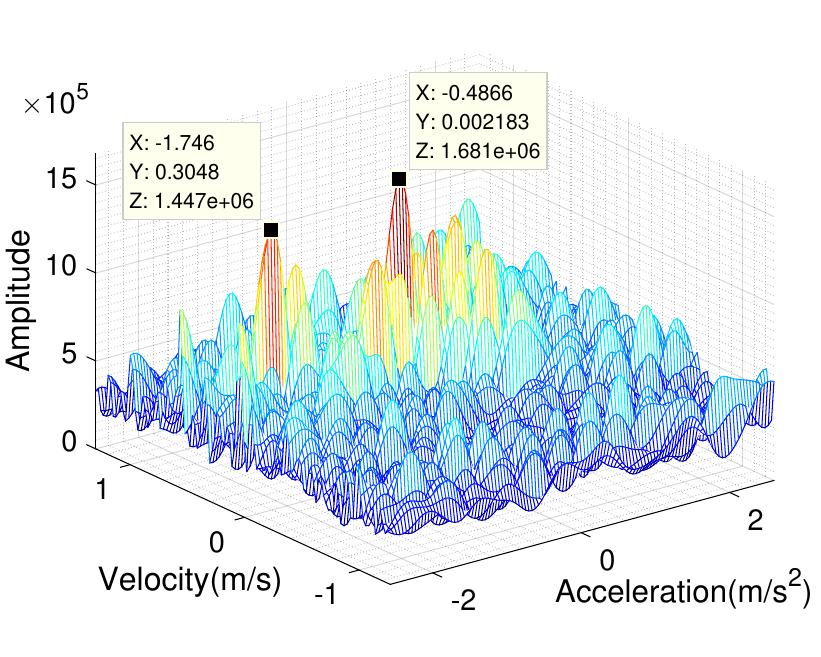}
%\caption{fig2}
\end{minipage}
} \\
\vspace{-0.4cm}
\subfigure[ ]{
\begin{minipage}[t]{0.3\linewidth}
\centering
\includegraphics[width=4.8cm]{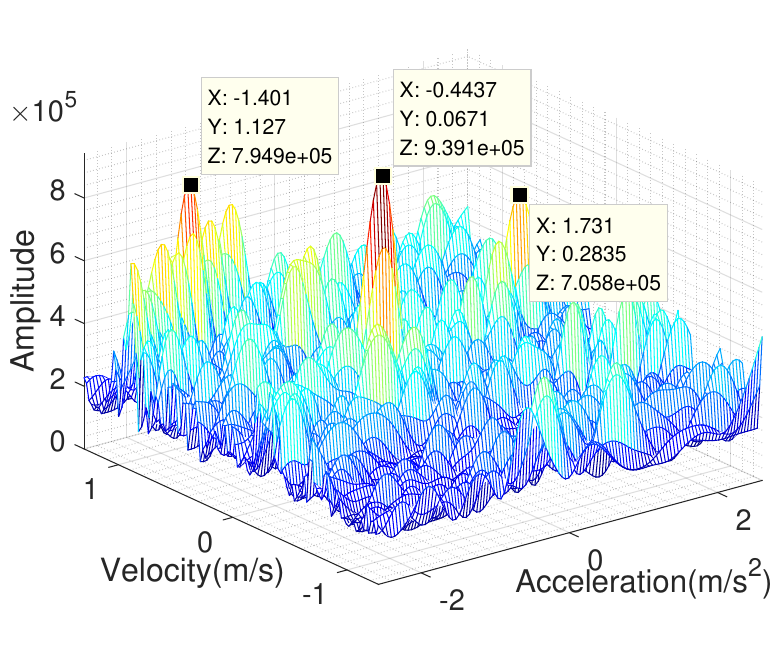}
%\caption{fig1}
\end{minipage}%
}%
\subfigure[ ]{
\begin{minipage}[t]{0.35\linewidth}
\centering
\includegraphics[width=5.3cm]{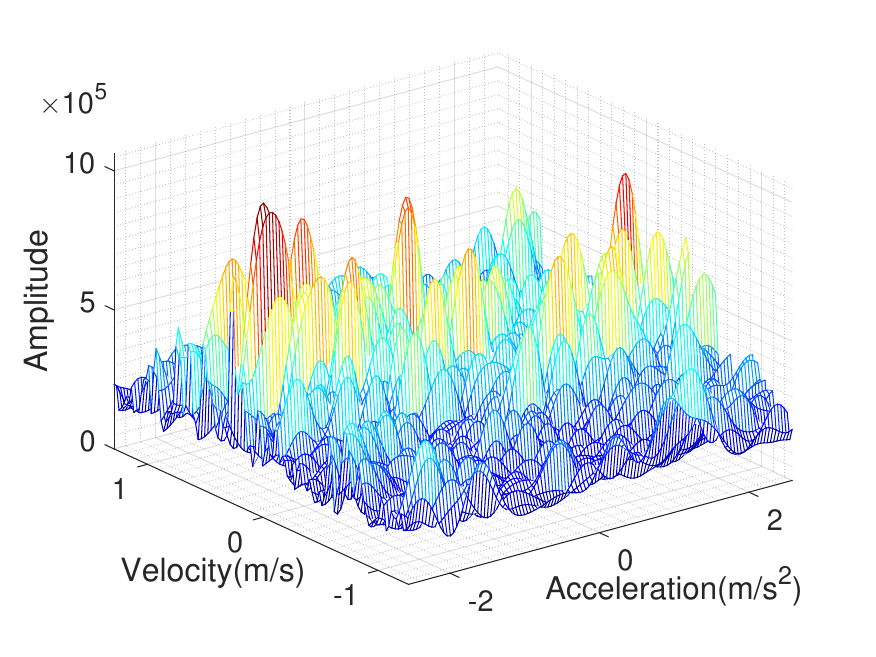}
%\caption{fig2}
\end{minipage}%
}%
\subfigure[ ]{
\begin{minipage}[t]{0.3\linewidth}
\centering
\includegraphics[width=5.3cm]{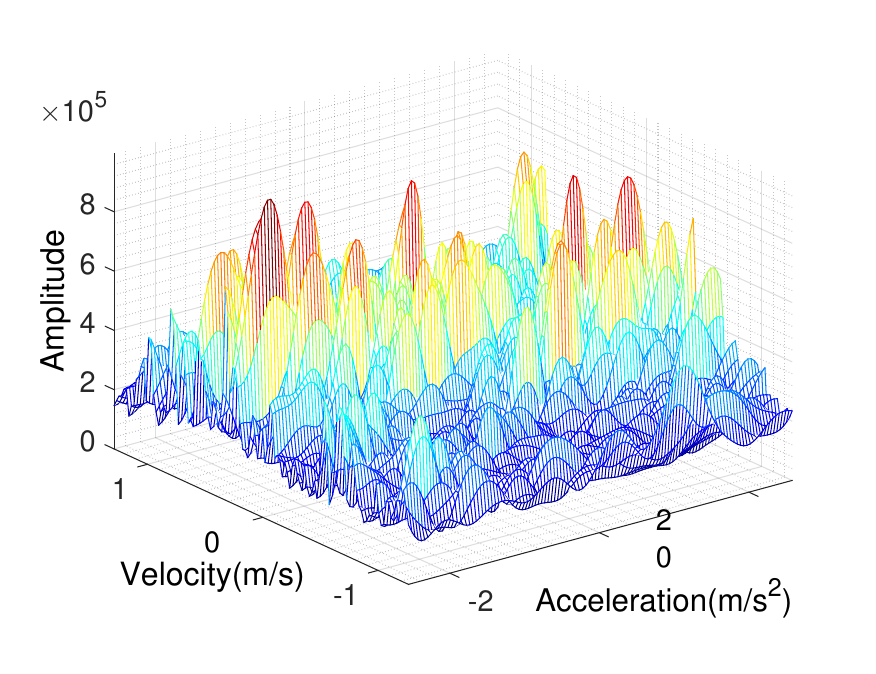}
%\caption{fig2}
\end{minipage}
}%
\centering
\caption{The parameter estimation results for different cases. The figure (a) and (b) are the estimation results for the single-target case without and with the elimination of direct and static reflection paths, respectively. The figures (c), (d), (e), and (f) show the estimation results for cases with two, three, four, and five moving people, respectively, after removing direct and static reflection paths.}
%\vspace{-0.8cm}
\label{SPCT}
\end{figure*}

First, two dimensional V-A plane generated by the proposed DP-AcE in different cases are presented in Fig.~\ref{SPCT}. As can be seen from Fig.~\ref{SPCT} (a), when the direct paths and the static reflection paths are not removed, only one peak with velocity and acceleration close to zero can be observed on the plane. This peak, resulting from the superposition of the direct path and the static reflection paths, is stronger than the DP introduced by moving human, making the DP difficult to identify.

After eliminating the direct and the static reflection paths, from Fig.~\ref{SPCT} (b), we observe that a peak characterized by strong energy and non-zero velocity and acceleration appears on the plane when a single human moves in the monitored area. As the number of moving humans increases to two and three, we observe a corresponding increase in the number of strong peaks on the plane, indicating that the proposed algorithm can effectively detect DPs and estimate the velocity and acceleration of moving humans. For the existing systems, this is challenging to achieve. However, when the number of moving humans increases further, to four and five, as shown in Figs.~\ref{SPCT} (e) and (f), the cross component and the mutual interference among individuals introduce greater noise, making the identification of the DPs introduced by the moving human targets more challenging.
% %\vspace{-0.5cm}
% \begin{figure*}[htbp]
% %\vspace{-.1cm}
% \centering
% %\hspace{-.5cm}
% \subfigure[]{
% \begin{minipage}[t]{0.2425\linewidth}
% \centering
% \includegraphics[width=4cm]{Image/WO_elimination.pdf}
% %\caption{fig1}
% \end{minipage}%
% }%
% \subfigure[]{
% \begin{minipage}[t]{0.2415\linewidth}
% \centering
% \includegraphics[width=4cm]{Image/SPCT_onehuman.pdf}
% %\caption{fig2}
% \end{minipage}%
% }%
% \subfigure[ ]{
% \begin{minipage}[t]{0.2345\linewidth}
% \centering
% \includegraphics[width=4cm]{Image/SPCT_TWOHUMAN2.pdf}
% %\caption{fig2}
% \end{minipage}
% }%
% \subfigure[]{
% \begin{minipage}[t]{0.245\linewidth}
% \centering
% \includegraphics[width=4cm]{Image/TD_SPC_positive_Trhuman.pdf}
% %\caption{fig2}
% \end{minipage}
% }%
% \centering
% \caption{The parameter estimation results for different cases. The figure (a) and (b) are the estimation result of the case where single moving human exist without and with static propagation paths elimination, respectively. The (c) and (d) show the estimation results with static propagation paths elimination for two and three moving people, respectively.}
% \label{F6}
% \end{figure*}

\begin{figure*}[htbp]
\centering
\includegraphics[height=6.5cm]{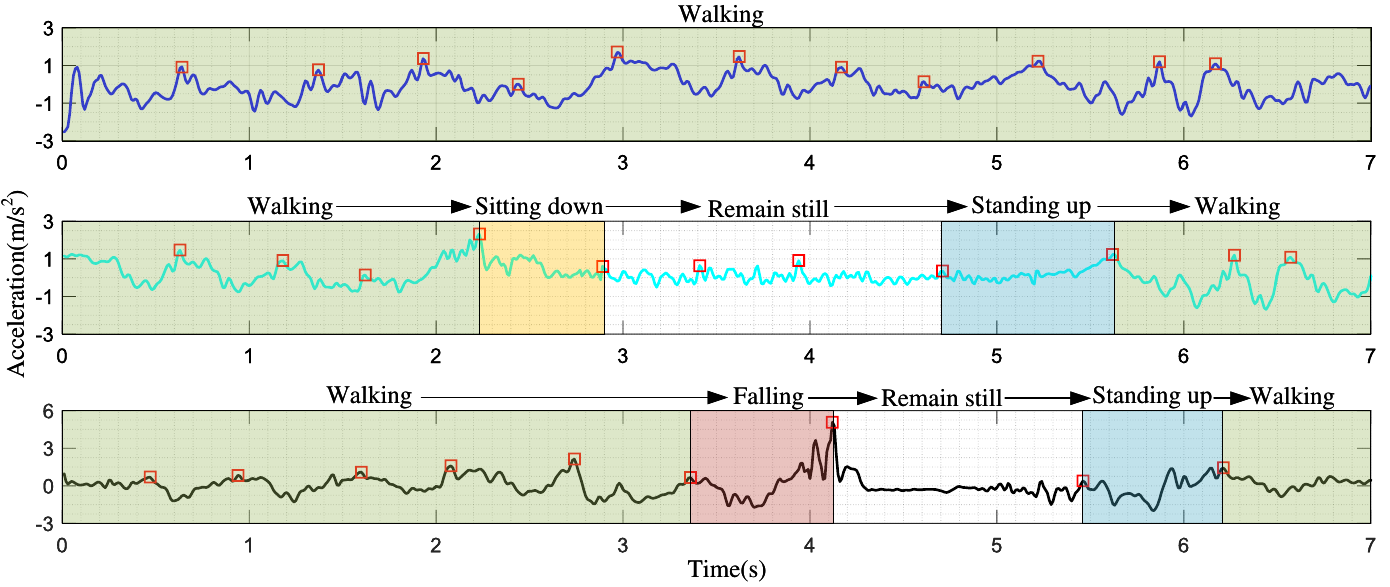}
%\vspace{-0.8cm}
\caption{The pattern of how acceleration varies over time for different actions.}
\label{Trend}
%\vspace{-0.3cm}
\end{figure*}

In addition to the estimation result analysis, the acceleration traces corresponding to different actions after filtering and segmentation are presented in Fig.~\ref{Trend}, where the detected local peaks are marked by red boxes. The figures reveal several significant observations. First, the acceleration during human walking follows a similar pattern to a sine wave, which is consistent with the results of existing research~\cite{umberger2010stance}, confirming the effectiveness of DP-AcE. Second, sitting down and standing up do not produce significant acceleration fluctuations, which is different from falling. The reason for this is that these two actions primarily trigger acceleration changes in the vertical direction, with little effect on the DPLC. Third, falling results in more distinct acceleration changes than those mentioned above, due to the sudden shift from an upright to a lying down position, providing a solid foundation for its detection.
%\vspace{-0.2cm}

\begin{figure*}[htbp]
  \centering
  \includegraphics[height=5.1cm]{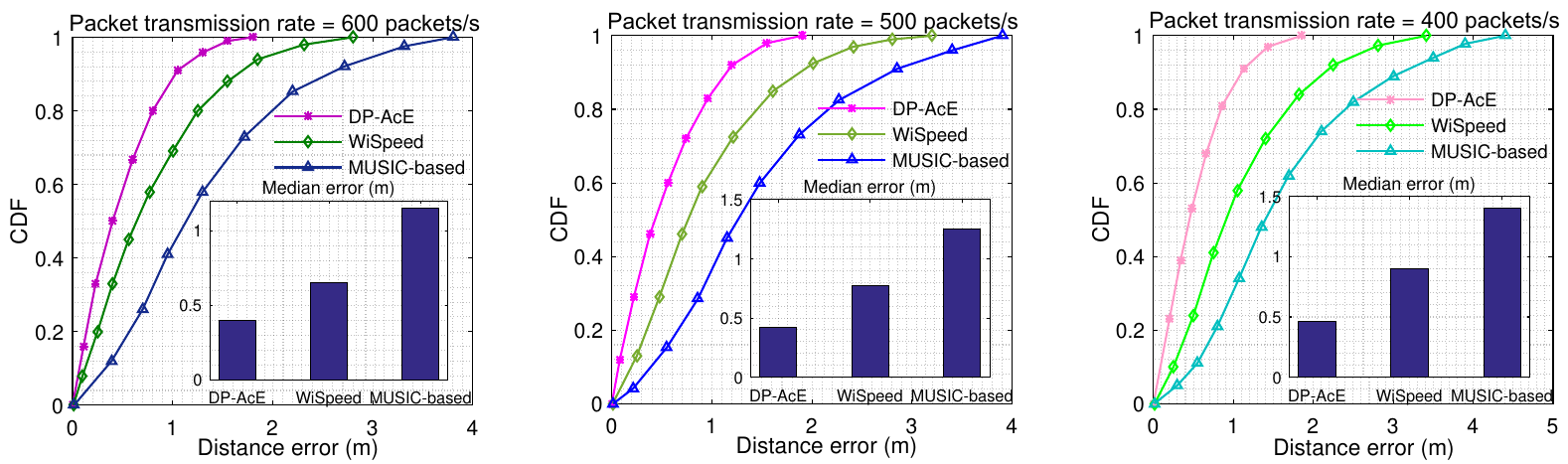} %1.png
   %\vspace{-1cm}
  \caption{Comparison of acceleration estimation accuracy of different algorithms under different packet transmission rate.}
  \label{AcCDF} %
  %\vspace{-0.5cm}
\end{figure*}

Through the integral operation, the estimated parameters can be converted into the walking distance. Taking the walking distance of 9.6 meters as a standard, here, the acceleration estimation accuracy of different algorithms is compared in the form of distance error. The results are presented in Fig.~\ref{AcCDF} via the cumulative distribution function (CDF) and the bar chart that describes the median error. Concretely, with a transmission rate of 600 packets/s, the DP-AcE's median estimation percentage error is about 4.38\% (about 0.42 m), which is better than the 8.02\% (about 0.77 m) of WiSpeed~\cite{zhang2018wispeed} and the 11.98\% (about 1.15 m) of the MUSIC-based method~\cite{li2017indotrack}, as shown by the first picture in Fig.~\ref{AcCDF}. This can be attributed to the fact that DP-AcE constructs a more accurate signal model and directly estimates the acceleration based on this model. Furthermore, we adjust the packet transmission rate to study its effect on the acceleration estimation accuracy. The results indicate that DP-AcE is able to maintain performance even when packet transmission rate is reduced, whereas the other two algorithms suffer from performance degradation to varying degrees, as shown in the second and third figures in Fig.~\ref{AcCDF}. This is because, even after a decrease in packet transmission rate, the transmission rate still exceeds twice the frequency of acceleration changes, enabling DP-AcE to estimate acceleration accurately.
%\vspace{-0.5cm}
%\vspace{-0.8cm}
\subsection{Fall detection performance analysis}
\begin{figure*}[htbp]
%\vspace{-.1cm}
\centering
%\hspace{-.5cm}
\subfigure[ ]{
\begin{minipage}[t]{0.2425\linewidth}
\centering
\includegraphics[width=4.5cm]{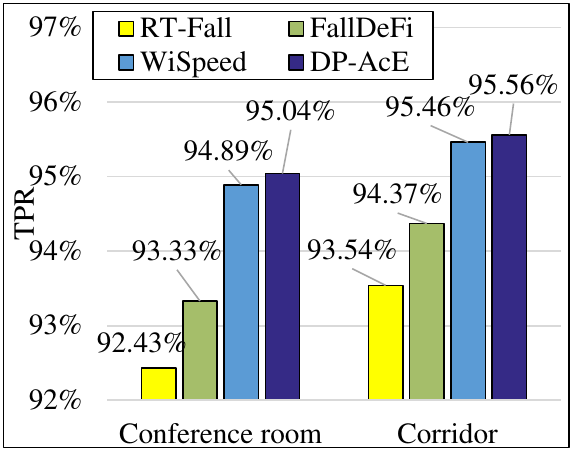}
%\caption{fig1}
\end{minipage}%
}%
\subfigure[ ]{
\begin{minipage}[t]{0.2415\linewidth}
\centering
\includegraphics[width=4.5cm]{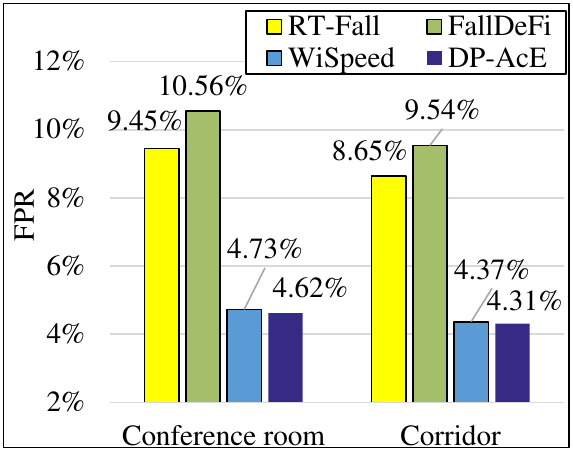}
%\caption{fig2}
\end{minipage}%
}%
\subfigure[ ]{
\begin{minipage}[t]{0.2345\linewidth}
\centering
\includegraphics[width=4.5cm]{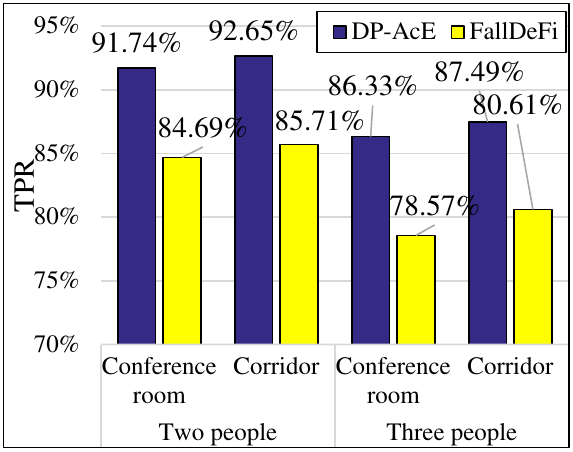}
%\caption{fig2}
\end{minipage}
}%
\subfigure[ ]{
\begin{minipage}[t]{0.245\linewidth}
\centering
\includegraphics[width=4.5cm]{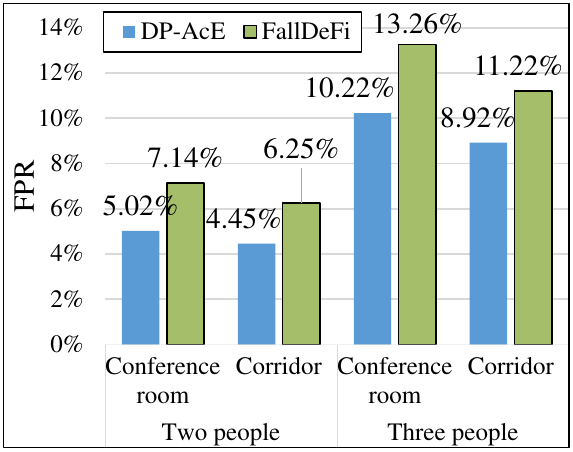}
%\caption{fig2}
\end{minipage}
}%
\centering
\caption{The fall detection performance comparison. Figures (a) and (b) illustrate the system's TPR and FPR when a single target appears in the monitored area, while figures (c) and (d) depict the TPR and FPR when multiple targets are present in the monitored area.}
\label{FD-PerF}
\end{figure*}

Following the parameter estimation performance analysis, the fall detection performance evaluation is carried out via the TPR and FPR comparison with RT-Fall, FallDeFi, and WiSpeed. First, in the single-target cases, the TPR and FPR of the different systems under the two experimental scenarios are presented in Figs.~\ref{FD-PerF} (a) and (b), respectively. The results show that in the conference room, DP-AcE's TPR can reach 95.04\%. However, under the same configuration, WiSpeed~\cite{zhang2018wispeed} achieves a TPR of 94.89\%, FallDeFi~\cite{palipana2018falldefi} realizes a TPR of 93.33\%, and RT-Fall~\cite{wang2016rt} is able to reach 92.43\%. These results indicate that our system can detect falls effectively, performing better than other methods. Furthermore, DP-AcE is less likely to produce false alarms. Rigorously, the DP-AcE's FPR can reach about 4.62\%, lower than WiSpeed's 4.73\%, RT-Fall's 9.45\%, and FallDeFi's 10.56\%. Compared to the conference room, the TPR of DP-AcE, WiSpeed, FallDeFi, and RT-Fall is increased by 0.52\%, 0.57\%, 1.04\%, and 1.11\%, respectively, in the corridor scenario. Moreover, in the corridor scenario, the FPR of each system is reduced to a different degree, but that of DP-AcE is still slightly lower than other systems. The above results not only demonstrate the effectiveness but also reveal a certain degree of generalization of the proposed DP-AcE, as the classifier, trained by the data collected in the conference room, performs well in the corridor scenario.

%\begin{figure*}
%	\centering
%	\subfigure[The detection accuracy.]{ \begin{minipage}{7cm}\includegraphics[width=\textwidth]{Image/Sg_DA2.pdf}
%		\end{minipage}
%	}
%	\subfigure[The false alarm rate.]{
%		\begin{minipage}{7cm}
%		\includegraphics[width=\textwidth]{Image/Sg_FAR2.pdf}
%		\end{minipage}
%	}
%\caption{The fall detection performance comparison in the case of the single human.}
%\label{F9}
%\end{figure*}
% \begin{figure*}[htbp]
% \label{PAAbefore}
% %\vspace{-.1cm}
% \centering
% %\hspace{-.5cm}
% \subfigure[The detection TPR.]{
% \begin{minipage}[t]{0.48\linewidth}
% \centering
% \includegraphics[width=8cm]{Image/Sg_DA4.pdf}
% %\caption{fig1}
% \end{minipage}%
% }%
% \subfigure[The detection FPR.]{
% \begin{minipage}[t]{0.52\linewidth}
% \centering
% \includegraphics[width=8cm]{Image/Sg_FAR4.pdf}
% %\caption{fig2}
% \end{minipage}%
% }%
% \centering
% \caption{The fall detection performance comparison in the case of the single human target.}
% \label{FDComp}
% %\vspace{-0.7cm}
% \end{figure*}
% Second, in the case of multiple human targets, the TPR and FPR of DP-AcE under both test scenarios are given in Figs.~\ref{FDComp} (a) and (b), respectively. As can be seen, the detection performance of the system decreases when more humans are present. Specifically, With the appearance of two human targets, the TPR of DP-AcE in the conference room and corridor drops to 91.74\% and 92.65\%, respectively. Meanwhile, the FPR increases to 5.02\% and 4.45\%. When the number of human targets reaches three, the TPR of DP-AcE in the conference room and corridor is further decreased to 86.33\% and 87.49\%, respectively, and the FPR increases to 10.22\% and 8.92\%. The main reason for these results is that the increasing number of human targets introduces more noise, which affects the acceleration estimation accuracy and eventually leads to a decline in detection performance. These experimental results validate the effectiveness of the DP-AcE based fall detection in multi-target scenarios, further highlighting its advantages over existing methods.
Second, in the case of multiple human targets, the TPR and FPR of DP-AcE are given in Figs.~\ref{FD-PerF} (a) and (b), respectively. Here, we only compare DP-AcE with FallDeFi, as it is capable of detecting falls in cases with more than one target~\cite{palipana2018falldefi}. As can be seen, the detection performance of the both systems decreases when more targets appear. Specifically, with the appearance of two human targets, the TPR of DP-AcE in the conference room and corridor drops to 91.74\% and 92.65\%, respectively, but still better than FallDeFi's 84.69\% and 85.71\%. Meanwhile, the FPR of DP-AcE increases to 5.02\% and 4.45\%, which are lower than FallDeFi's 7.14\% and 6.25\%. When the number of human targets reaches three, the performance of both systems further declines. For instance, in the corridor, the TPR and FPR of DP-AcE further deteriorate to 87.49\% and 8.92\%, respectively, while FallDeFi's worsen to 80.61\% and 11.2\%. The main reason for these results is that the increasing number of human targets introduces more noise, which affects the signal feature extraction and eventually leads to a decline in detection performance. Nevertheless, DP-AcE still performs better overall compared to FallDeFi, confirming that our system works well in multi-target cases and highlighting its strengths over the existing method.

\begin{figure*}[htbp]
%\vspace{-.1cm}
\centering
%\hspace{-.5cm}
\subfigure[ ]{
\begin{minipage}[t]{0.2425\linewidth}
\centering
\includegraphics[width=4.5cm]{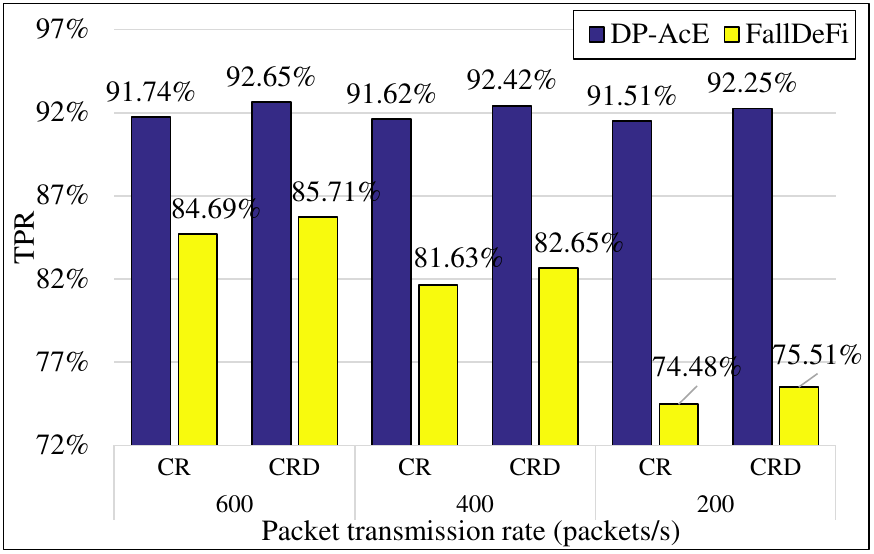}
%\caption{fig1}
\end{minipage}%
}%
\subfigure[ ]{
\begin{minipage}[t]{0.2415\linewidth}
\centering
\includegraphics[width=4.5cm]{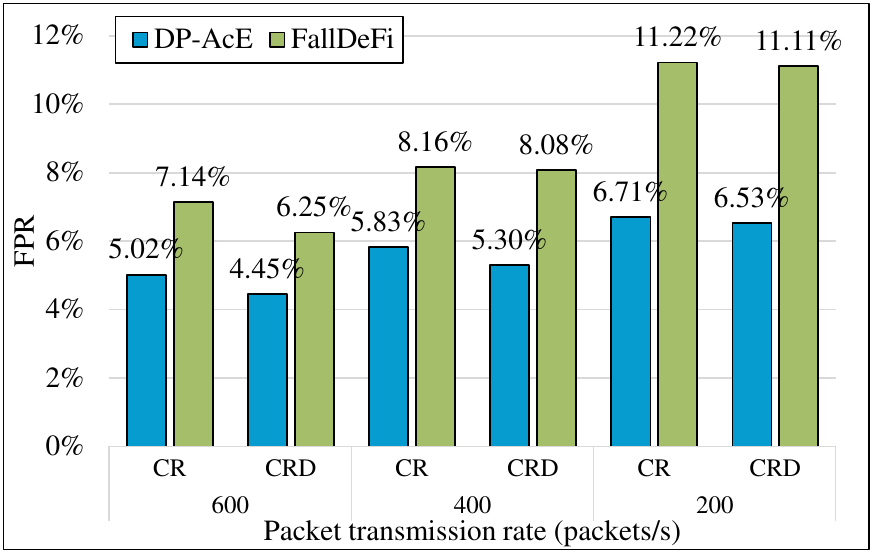}
%\caption{fig2}
\end{minipage}%
}%
\subfigure[ ]{
\begin{minipage}[t]{0.2345\linewidth}
\centering
\includegraphics[width=4.5cm]{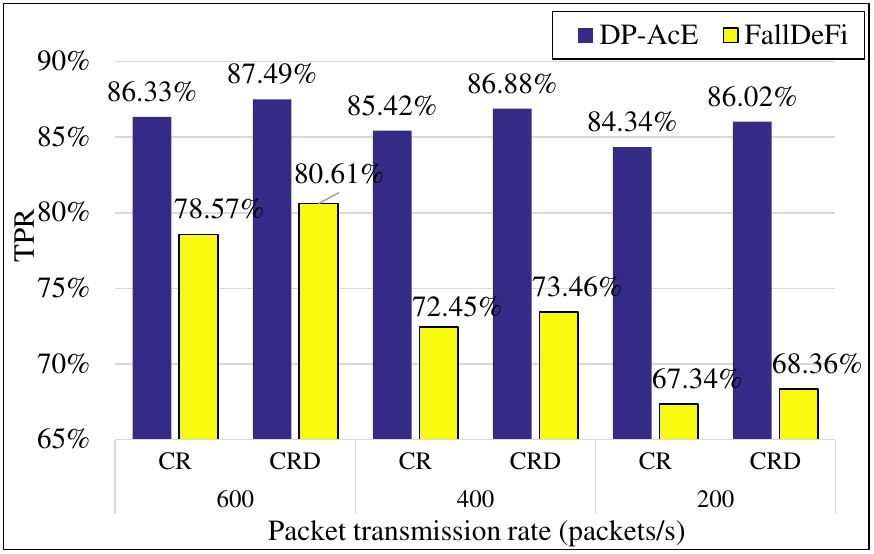}
%\caption{fig2}
\end{minipage}
}%
\subfigure[ ]{
\begin{minipage}[t]{0.245\linewidth}
\centering
\includegraphics[width=4.5cm]{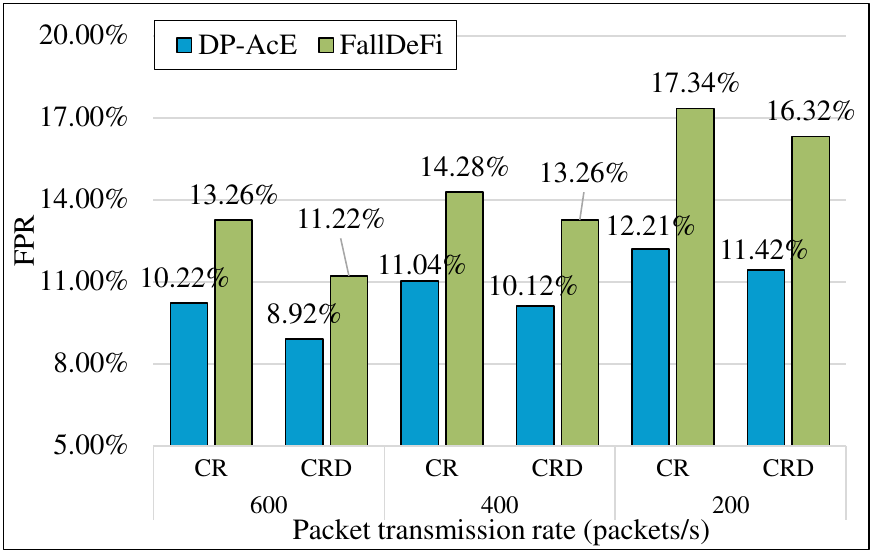}
%\caption{fig2}
\end{minipage}
}%
\centering
\caption{The impact of packet transmission rate on the detection performance when multiple targets appear in the monitored area. Figures (a) and (b) illustrate the system's TPR and FPR when two targets appear in the monitored area, while figures (c) and (d) depict the TPR and FPR when three targets are present in the monitored area. Here, the CR means conference room, and CRD stands for corridor.}
\label{PTR}
\end{figure*}

The impact of packet transmission rate on the detection performance is investigated in the case of two and three human targets, and the results are presented in Fig.~\ref{PTR}. For instance, in a conference room with two targets, DP-AcE's TPR is around 91.74\%, 91.62\%, and 91.51\% for packet transmission rates of 600, 400, and 200 packets/s, respectively. In comparison, FallDeFi's TPR is lower at 84.69\%, 81.63\%, and 74.48\% for the same configurations. Regarding the FPR, DP-AcE records approximately 5.02\%, 5.83\%, and 6.71\%, which are lower than FallDeFi's FPRs of 7.14\%, 8.16\%, and 11.22\%. In the corridor scenario, a similar pattern can be observed, i.e., as the packet transmission rate decreases, the system's TPR decreases and the FPR gradually increases. When three targets appear in the monitored area, the decline in TPR and increase in FPR become more noticeable. For example, in the corridor, when the packet transmission rate drops from 600 to 200 packets/s, the DP-AcE's TPR falls from 87.49\% to 86.02\%, while the FPR goes up from 8.92\% to 11.42\%. For the FallDeFi, the TPR declines from 80.61\% to 68.36\%, while the FPR rises from 11.22\% to 16.32\%. Although the overall performance is impacted, the proposed system can still effectively detect falls.

%\vspace{-0.5cm}
\begin{figure*}[htbp]
%\vspace{-.1cm}
\centering
%\hspace{-.5cm}
\subfigure[]{
\begin{minipage}[t]{0.2425\linewidth}
\centering
\includegraphics[width=4.5cm]{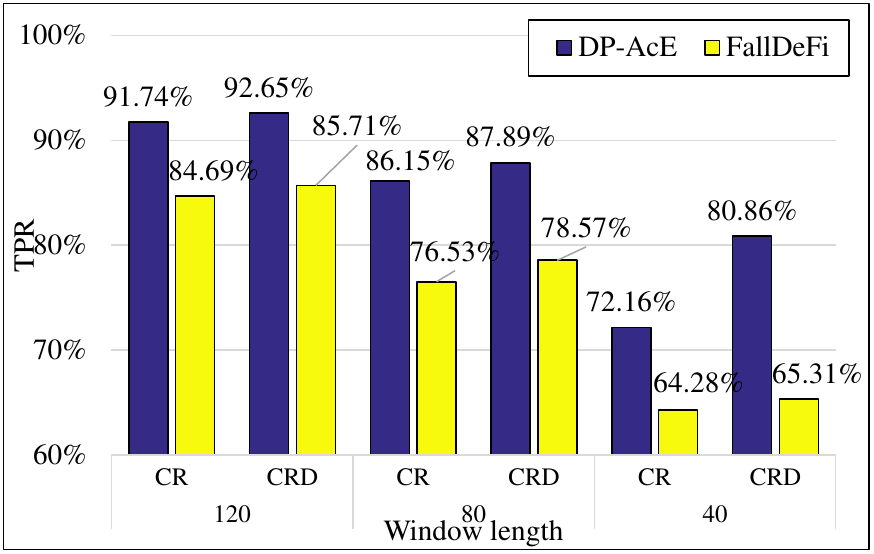}
%\caption{fig1}
\end{minipage}%
}%
\subfigure[]{
\begin{minipage}[t]{0.2415\linewidth}
\centering
\includegraphics[width=4.5cm]{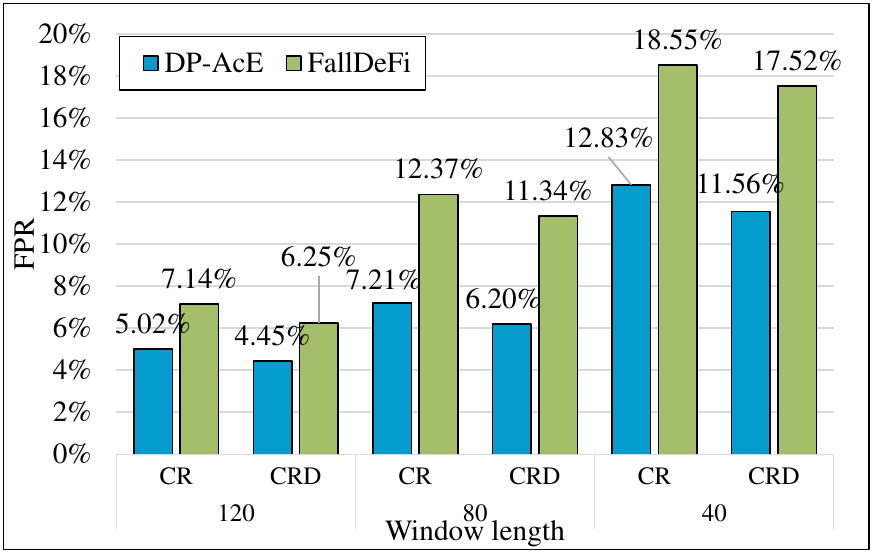}
%\caption{fig2}
\end{minipage}%
}%
\subfigure[ ]{
\begin{minipage}[t]{0.2345\linewidth}
\centering
\includegraphics[width=4.5cm]{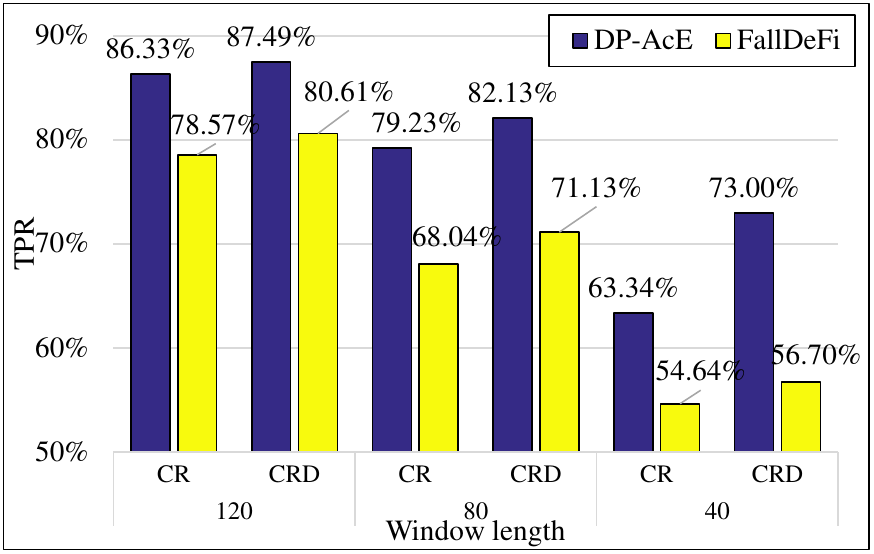}
%\caption{fig2}
\end{minipage}
}%
\subfigure[]{
\begin{minipage}[t]{0.245\linewidth}
\centering
\includegraphics[width=4.5cm]{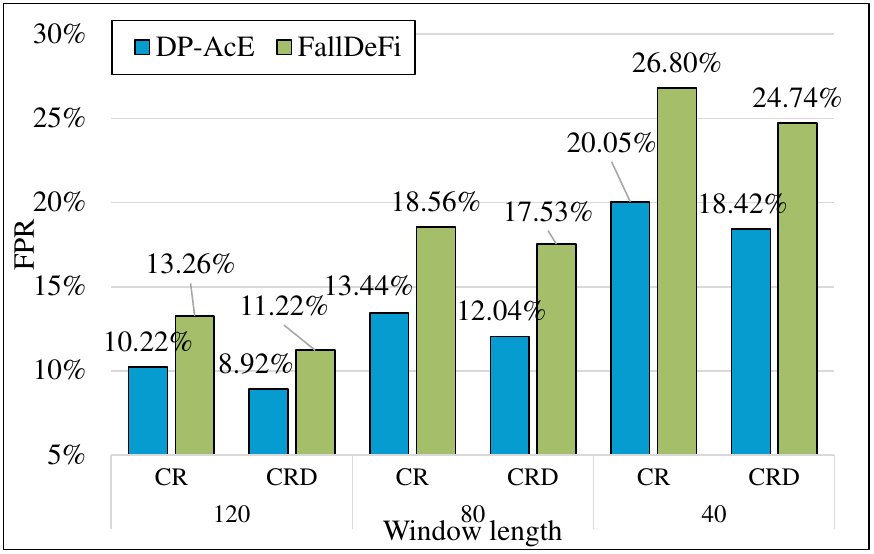}
%\caption{fig2}
\end{minipage}
}%
\centering
\caption{The impact of window length on the detection performance when multiple targets appear in the monitored area. Figures (a) and (b) illustrate the system's TPR and FPR when two targets appear in the monitored area, while figures (c) and (d) depict the TPR and FPR when three targets appear. Here, the CR means conference room, and CRD stands for corridor.}
\label{WL}
\end{figure*}
Compared to the packet transmission rate, the window length has a greater effect on the detection performance, as the results in Fig.~\ref{WL} show. For instance, when two moving targets appear in the conference room and the window length is set to 120, 80, and 40, the TPR of DP-AcE achieves approximately 91.74\%, 86.15\%, and 72.16\%, respectively. In contrast, FallDeFi exhibits lower TPR values of 84.69\%, 76.53\%, and 64.28\% under the same settings. Regarding the FPR, DP-AcE achieves values of approximately 5.02\%, 7.21\%, and 12.83\%, lower than that of FallDeFi, which are 7.14\%, 12.37\%, and 18.55\%. Similarly, in the corridor scenario, the performance of both systems degrades as the window length gets shorter. When three targets appear in the monitored area, such degradation becomes more obvious. For example, in the corridor, we observe that the TPR of DP-AcE decrease from 87.49\% to 73.00\%, and FPR increase from 8.92\% to 18.42\%, when the window length drops from 120 to 40. Comparatively, the performance degradation of FallDeFi is more significant, with its TPR falling from 80.61\% to 56.70\% and FPR increasing from 11.22\% to 24.74\%.

This is reasonable, since shortening the window length reduces the amount of CSI measurements involved in the estimation. According to the basic principles of FFT, such a reduction lowers the signal-to-noise ratio (SNR), as well as the resolution of the V-A plane. Figure 14 shows the V-A planes generated by the proposed algorithm with different window lengths. The results demonstrate that the resolution of the plane and the strength of peaks significantly decreases with the reduction of the window length. This degradation negatively affects the system's ability to estimate the acceleration and, ultimately, impacts the fall detection performance. While affected, DP-AcE still performs better than FallDeFi, demonstrating its stronger adaptability to multi-target cases.

\begin{figure*}[htbp]
%\vspace{-.1cm}
%\vspace{-0.5cm}
\centering
%\hspace{-.5cm}
\subfigure[The window length $W=140$.]{
\begin{minipage}[t]{0.3\linewidth}
\centering
\includegraphics[width=4.8cm]{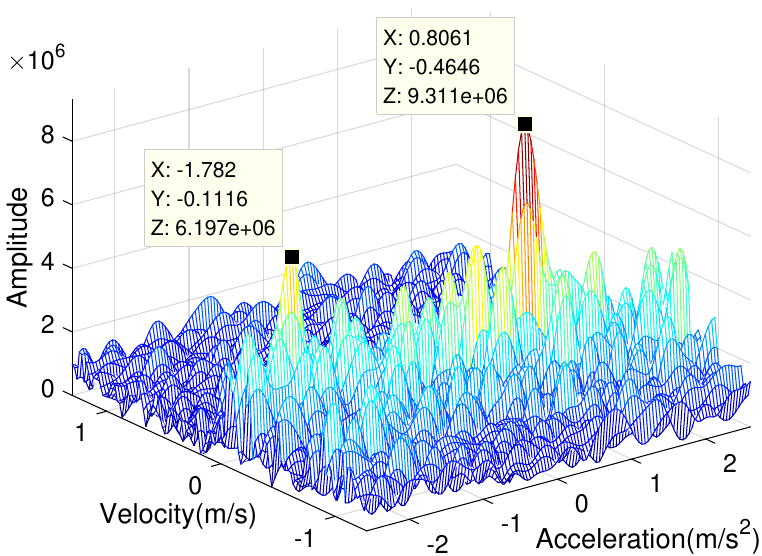}
%\caption{fig1}
\end{minipage}%
}%
\subfigure[The window length $W=120$.]{
\begin{minipage}[t]{0.35\linewidth}
\centering
\includegraphics[width=4.8cm]{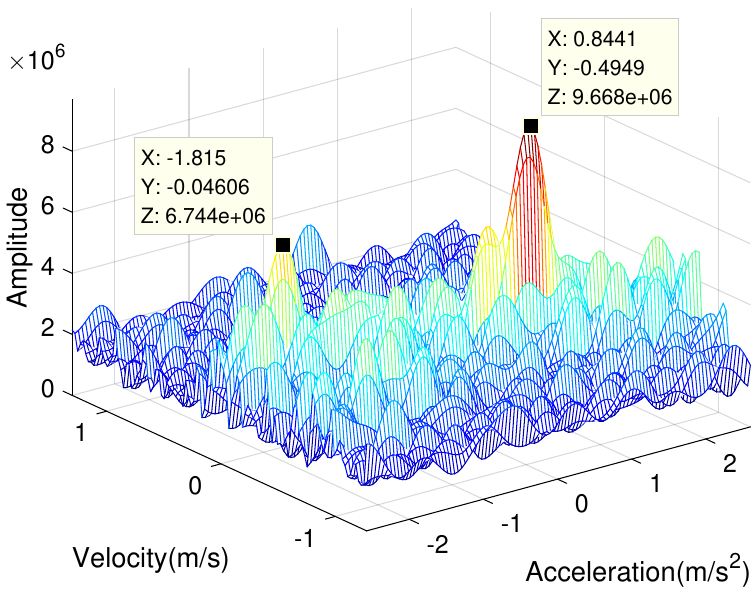}
%\caption{fig2}
\end{minipage}%
}%
\subfigure[The window length $W=100$.]{
\begin{minipage}[t]{0.3\linewidth}
\centering
\includegraphics[width=4.8cm]{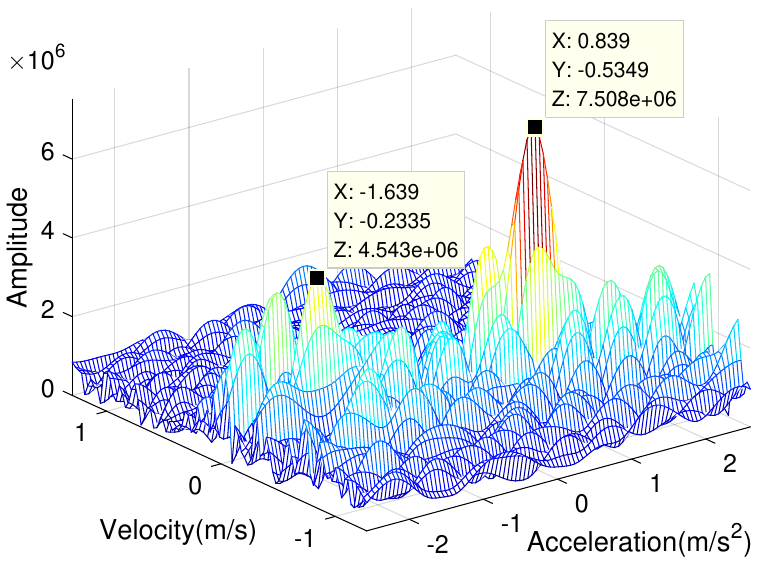}
%\caption{fig2}
\end{minipage}
} \\
\subfigure[The window length $W=80$.]{
\begin{minipage}[t]{0.3\linewidth}
\centering
\includegraphics[width=4.8cm]{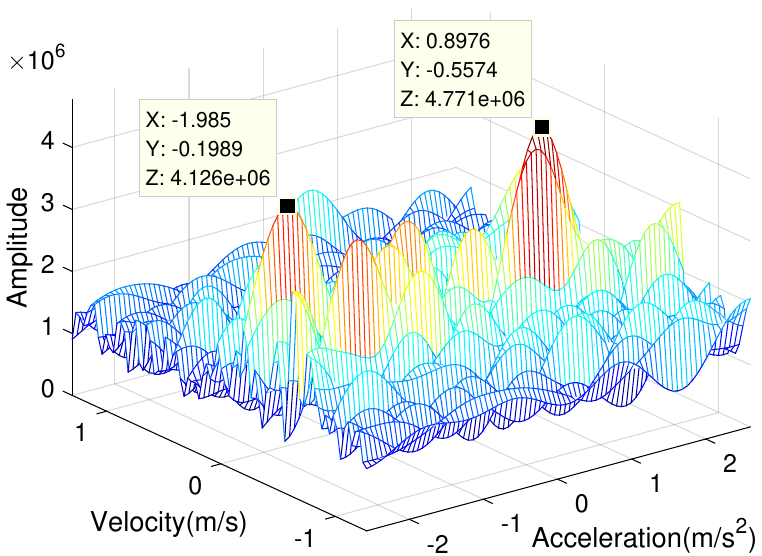}
%\caption{fig1}
\end{minipage}%
}%
\subfigure[The window length $W=60$.]{
\begin{minipage}[t]{0.35\linewidth}
\centering
\includegraphics[width=4.8cm]{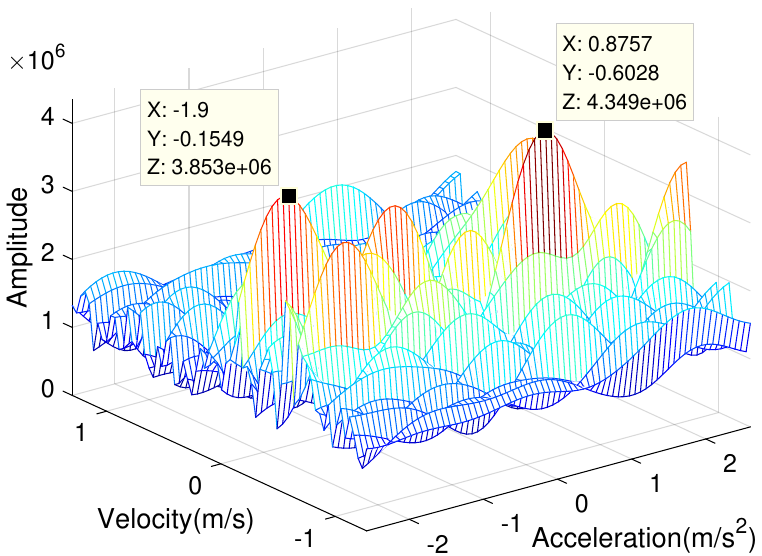}
%\caption{fig2}
\end{minipage}%
}%
\subfigure[The window length $W=40$.]{
\begin{minipage}[t]{0.3\linewidth}
\centering
\includegraphics[width=4.8cm]{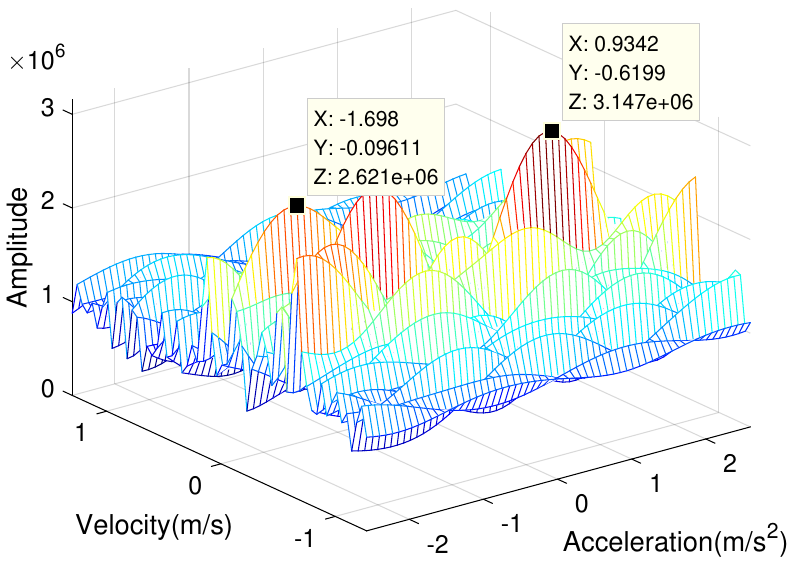}
%\caption{fig2}
\end{minipage}
}%
\centering
\caption{ The impact of the length of sliding window on the velocity and acceleration estimation. As shown in the figures, a decrease in window length results in a reduction of the number of CSI packets involved in the estimation. This results in a decline in resolution and SNR, leading to a gradual increase in estimation errors.}
%\vspace{-0.8cm}
\label{F13}
\end{figure*}
%\vspace{-0.4cm}
\subsection{Discussion}
In the above experiments, we evaluate the proposed algorithm from the aspects of acceleration estimation accuracy and fall detection, and analyze the influence of parameters such as packet transmission rate on the algorithm. From the results, we can summarize the following key points:
\begin{itemize}
        \item The proposed model is more accurate than existing models, and the proposed estimation algorithm can effectively estimate the acceleration of DPLC based on the constructed model, with a higher accuracy than those of the existing algorithms.

       \item The proposed method can distinguish the velocity and acceleration of different DPs, offering advantages in sensing tasks that require simultaneous detection and analysis of multiple targets, as validated in fall detection.

        \item The estimation accuracy of the proposed method is related to the amount of data involved in estimation. Therefore, given a fixed packet transmission rate, it is recommended to increase the length of sliding window as much as possible to enhance the performance.

        \item Since the acceleration is environment-independent, utilizing it to construct sensing systems yields superior generalization performance. For instance, a fall detection classifier trained with data gathered from a conference room retains its effectiveness in the corridor.
   \end{itemize}
 Besides the above key findings, the proposed DP-AcE also has the following limitations:
 \begin{itemize}
        \item Our model considers the velocity and acceleration of DPLC, but for situations with more complex dynamic characteristics, it is necessary to consider more factors such as changes in acceleration, to construct a more comprehensive model.

       \item The estimation performance is related to the data amount involved in each estimation. Therefore, to ensure optimal sensing performance and real-time capabilities in practical scenarios, it is recommended to avoid excessively low packet transmission rate.

        \item While our algorithm can distinguish multiple targets, the presence of more targets can increase the noise, resulting in a reduction in estimation accuracy. This issue can be alleviated by augmenting the packet transmission rate and the length of sliding window.

   \end{itemize}

\section{CONCLUSION }
In this paper, we propose DP-AcE, which can directly estimate the acceleration of DPLC based on the CSI extracted from wireless communication signals. Furthermore, we illustrate the importance of acceleration in wireless sensing applications through fall detection. Specifically, we establish a model that explains the relationship between the phase difference of two consecutive CSI measurements and the velocity and acceleration of DPLC. Subsequently, a novel algorithm is introduced to estimate the acceleration of DPLC utilizing the established model. Using the estimated acceleration, we extract a series of statistical features and train an SVM to achieve human fall detection. The experimental results reveal that, using distance as the metric, DP-AcE yields a median estimation percentage error of 4.38\% for acceleration estimation, outperforms existing methods. Furthermore, when more than one targets appears in the monitored area, DP-AcE based fall detection realizes an average true positive rate of 89.56\% and a false positive rate of 11.78\%. These results highlight the advantages of our method in multi-target scenarios and demonstrate its importance in enhancing indoor wireless sensing capabilities. In the future, we will focus on eliminating cross-component and mutual interference to improve our algorithm and study how estimated acceleration can enhance other sensing systems, including target tracking and behavior recognition.

%\bibliographystyle{IEEEtran}
% \bibliography{references.bib}
%\bibliography{Ref.bib}
%1

%\footnotesize
%\setlength{\bibsep}{1ex}
\bibliographystyle{IEEEtran}
%\begin{spacing}{1.2}
 \bibliography{Ref}
%\end{spacing}

% biography section
%
% If you have an EPS/PDF photo (graphicx package needed) extra braces are
% needed around the contents of the optional argument to biography to prevent
% the LaTeX parser from getting confused when it sees the complicated
% \includegraphics command within an optional argument. (You could create
% your own custom macro containing the \includegraphics command to make things
% simpler here.)
%\begin{IEEEbiography}[{\includegraphics[width=1in,height=1.25in,clip,keepaspectratio]{mshell}}]{Michael Shell}
% or if you just want to reserve a space for a photo:

%\begin{IEEEbiography}{Michael Shell}
%Biography text here.
%\end{IEEEbiography}

% if you will not have a photo at all:
%\begin{IEEEbiographynophoto}{John Doe}
%Biography text here.
%\end{IEEEbiographynophoto}

% insert where needed to balance the two columns on the last page with
% biographies
%\newpage

%\begin{IEEEbiographynophoto}{Jane Doe}
%Biography text here.
%\end{IEEEbiographynophoto}

% You can push biographies down or up by placing
% a \vfill before or after them. The appropriate
% use of \vfill depends on what kind of text is
% on the last page and whether or not the columns
% are being equalized.

%\vfill

% Can be used to pull up biographies so that the bottom of the last one
% is flush with the other column.
%\enlargethispage{-5in}

% that's all folks
\end{document}